\PassOptionsToPackage{unicode}{hyperref}
\PassOptionsToPackage{hyphens}{url}
\PassOptionsToPackage{dvipsnames,svgnames,x11names}{xcolor}
\documentclass[
  11pt,
]{article}
\usepackage{amsmath,amssymb}
\usepackage{iftex}
\ifPDFTeX
  \usepackage[T1]{fontenc}
  \usepackage[utf8]{inputenc}
  \usepackage{textcomp} 
\else 
  \usepackage{unicode-math} 
  \defaultfontfeatures{Scale=MatchLowercase}
  \defaultfontfeatures[\rmfamily]{Ligatures=TeX,Scale=1}
\fi
\usepackage{lmodern}
\ifPDFTeX\else
\fi
\IfFileExists{upquote.sty}{\usepackage{upquote}}{}
\IfFileExists{microtype.sty}{
  \usepackage[]{microtype}
  \UseMicrotypeSet[protrusion]{basicmath} 
}{}
\usepackage{xcolor}
\usepackage[margin=1in]{geometry}
\usepackage{graphicx}
\makeatletter
\def\maxwidth{\ifdim\Gin@nat@width>\linewidth\linewidth\else\Gin@nat@width\fi}
\def\maxheight{\ifdim\Gin@nat@height>\textheight\textheight\else\Gin@nat@height\fi}
\makeatother
\setkeys{Gin}{width=\maxwidth,height=\maxheight,keepaspectratio}
\makeatletter
\def\fps@figure{htbp}
\makeatother
\providecommand{\tightlist}{%
  \setlength{\itemsep}{0pt}\setlength{\parskip}{0pt}}
\newlength{\cslhangindent}
\newlength{\csllabelwidth}
\newlength{\cslentryspacingunit} 
\newenvironment{CSLReferences}[2] 
 {
  \setlength{\parindent}{0pt}
  \ifodd #1
  \let\oldpar\par
  \def\par{\hangindent=\cslhangindent\oldpar}
  \fi
  \setlength{\parskip}{#2\cslentryspacingunit}
 }%
 {}
\usepackage{calc}

\usepackage{accents}
\usepackage{amsmath}
\usepackage{bm}
\usepackage{booktabs}
\usepackage{braket}
\usepackage{caption}
\usepackage{dcolumn}
\usepackage{epigraph}
\usepackage{extarrows}
\usepackage{float}
\usepackage[bottom,flushmargin,hang,multiple]{footmisc}
\usepackage{graphicx}
\usepackage{hyperref}
\usepackage[export]{adjustbox}
\usepackage{libertine}
\usepackage[libertine]{newtxmath}
\usepackage[mathcal]{euscript}
\usepackage{mathrsfs}
\usepackage{multirow}
\usepackage{pdflscape}
\usepackage{quoting}
\usepackage{scrextend}
\usepackage{setspace}
\usepackage{soul}
\usepackage{subcaption}
\usepackage{tabularx}
\usepackage{titlesec}
\usepackage[utf8]{inputenc}
\ifLuaTeX
  \usepackage{selnolig}  
\fi
\IfFileExists{bookmark.sty}{\usepackage{bookmark}}{\usepackage{hyperref}}
\IfFileExists{xurl.sty}{\usepackage{xurl}}{} 
\hypersetup{
  pdftitle={CAVIAR: Categorical-Variable Embeddings for Accurate and Robust Inference},
  colorlinks=true,
  linkcolor={Maroon},
  filecolor={Maroon},
  citecolor={Blue},
  urlcolor={blue},
  pdfcreator={LaTeX via pandoc}}

\title{CAVIAR: Categorical-Variable Embeddings for Accurate and Robust
Inference}
\author{}
\date{\vspace{-2.5em}}

\begin{document}
\maketitle

\quotingsetup{indentfirst=false, leftmargin=2em, rightmargin=2em, vskip=1ex}
\begin{center}
\author
{Anirban Mukherjee,$^{1\ast}$ Hannah H. Chang$^{2}$\\
\medskip
\normalsize{$^{1}$Samuel Curtis Johnson Graduate School of Management, Cornell University,}\\
\normalsize{Sage Hall, Ithaca, NY 14850, USA}\\
\normalsize{$^{2}$Lee Kong Chian School of Business, Singapore Management University,}\\
\normalsize{50 Stamford Road, Singapore, 178899}\\
\smallskip
\normalsize{$^\ast$To whom correspondence should be addressed; E-mail: am253@cornell.edu.}\\
}
\end{center}
\medskip
\singlespacing

\begin{center} 
\noindent \textbf{Abstract}
\end{center}

\noindent Social science research often hinges on the relationship
between categorical variables and outcomes. We introduce CAVIAR, a novel
method for embedding categorical variables that assume values in a
high-dimensional ambient space but are sampled from an underlying
manifold. Our theoretical and numerical analyses outline challenges
posed by such categorical variables in causal inference. Specifically,
dynamically varying and sparse levels can lead to violations of the
Donsker conditions and a failure of the estimation functionals to
converge to a tight Gaussian process. Traditional approaches, including
the exclusion of rare categorical levels and principled variable
selection models like LASSO, fall short. CAVIAR embeds the data into a
lower-dimensional global coordinate system. The mapping can be derived
from both structured and unstructured data, and ensures stable and
robust estimates through dimensionality reduction. In a dataset of
direct-to-consumer apparel sales, we illustrate how high-dimensional
categorical variables, such as zip codes, can be succinctly represented,
facilitating inference and analysis.

\begin{center}\rule{0.5\linewidth}{0.5pt}\end{center}


\newpage
\doublespacing

\hypertarget{introduction}{%
\section{Introduction}\label{introduction}}

This paper addresses concerns related to the estimation of causal
econometric models in the social sciences in which categorical
variables, taking values in an ambient space and presumed to be sampled
from an underlying manifold, are mapped to dependent variables of
interest.

Distance within the manifold is considered a measure of similarity
relevant to the analysis. For instance, when analyzing a categorical
variable that describes color, distance may correspond to the light
spectrum, such that bottle green is further from maroon than from
turquoise. Similarly, when examining a spatial categorical variable,
distance may relate to physical proximity, indicating that a city is
closer to its suburb than to a distant location. When analyzing sound,
distance may be related to frequencies, implying that the sounds of a
viola and violin are more similar than those from a snare drum. In
classical methods, distinct elements of the set are accorded distinct
levels, such that each color, location, and instrument might have its
own level in corresponding categorical variables.

Categorical variables often present two complexities: large and
increasing cardinality with sample size (the categorical variable
assumes many distinct levels, with new levels added as new samples are
introduced) and sparsity (levels that correspond to only a few
observations). These conditions may occur when specificity is desired.
For instance, consider the categorization of religion---a typical
covariate in the social sciences
(\protect\hyperlink{ref-fox2001religion}{Fox 2001},
\protect\hyperlink{ref-woodberry2012missionary}{Woodberry 2012}). When
categorized broadly, a religion categorical variable may be developed to
include only a few levels (e.g., Christianity, Islam, etc.).

However, when more granularity is desired, the categorical variable may
encompass many distinct values
(\protect\hyperlink{ref-center2015future}{Center 2015},
\protect\hyperlink{ref-finke2005churching}{Finke and Stark 2005}). For
example, there may be thousands of denominations and sects, leading to a
system that is both characterized by large cardinality and marked by
sparsity, in the sense that, for many designations, a particular dataset
may contain only one or a few observations
(\protect\hyperlink{ref-smith1990classifying}{Smith 1990},
\protect\hyperlink{ref-steensland2000measure}{Steensland et al. 2000}).
Moreover, as the data grows, observations may be added that pertain to
more detailed sub-categories whose inclusion in the analysis as distinct
levels may be desirable
(\protect\hyperlink{ref-voas2013modernization}{Voas et al. 2013},
\protect\hyperlink{ref-woodberry2012measure}{Woodberry et al. 2012}).

These issues complicate causal inference. Typically, a fixed effects
model is specified whereby each level of the categorical variable is
accorded its own parameter, which is then estimated freely; the
traditional fixed-effects model is a non-parametric model. This
methodology is ubiquitous and has been used countless times in the
social sciences
(\protect\hyperlink{ref-wooldridge2010econometric}{Wooldridge 2010}). It
is robust and well-understood in cases where the categorical data
exhibits low cardinality and high density of occurrence, which are the
canonical scenarios for which it was originally proposed
(\protect\hyperlink{ref-suits1957use}{Suits 1957}).

In scenarios where the data exhibits high cardinality and sparsity, the
canonical fixed effects model can break down, leading to inaccurate and
imprecise inference. Here, we depart from existing studies by
considering cases of extreme sparsity where a categorical variable may
have thousands of possible levels conceptually (e.g., thousands of
possible religious subdivisions), only a fraction of which may be
represented in the data. In such cases, the categorical variable is
effectively increasing in levels and dimensionality with the inclusion
of more data. Additionally, our model maps onto situations where the
context is dynamic, and therefore, the categorical variable is
dynamic---over time, new category levels emerge (e.g., new religious
sects are formed) and old category levels become irrelevant (e.g.,
extant religious sects become dormant). These considerations challenge
prior econometric models, which were generally developed for scenarios
where a categorical variable reflects a few fixed dimensionality
categorical factors, such as might emerge from the use of a Likert scale
in a survey.

To combat this issue, researchers often resort to ad-hoc solutions, such
as collapsing rare levels into a `meta-level' or applying formal
regularization methods like the Least Absolute Shrinkage and Selection
Operator (LASSO) to the indicator variables of the levels
(\protect\hyperlink{ref-tutz2016regularized}{Tutz and Gertheiss 2016}).
However, both approaches can lead to inaccurate and imprecise inference;
this is because, in the fixed effects model, inference is only drawn
from observations that relate to a specific level of the categorical
variable. Consequently, the addition of levels to the categorical
variable---whether as they appear in the data (the traditional model),
when they occur a specified number of times (the ad-hoc solution), or
when the fixed effect coefficient is `sufficiently' large or the design
matrix is `sufficiently' non-singular (the regularized solution)---can
result in violations of the Donsker conditions. This, in turn,
eliminates the guarantee that the estimators, as functionals of the
empirical process, will converge to a Brownian bridge.

Addressing this research gap, we propose a method for the development of
a categorical variable embedding---an injunction from the
high-dimensional ambient space to a single global coordinate system of
lower dimensionality, where distance captures a notion of similarity in
the categorical levels, such that the location of each level can act as
a data feature in the causal model. Fixed effects in the original causal
model are recovered through their projection onto the lower-dimensional
space.

Our underlying research agenda speaks to the development of a data
pipeline that averts issues linked to categorical variable cardinality
and sparsity. In relatively straightforward cases, our method echoes the
intuition of replacing the fixed effects with a structural model of
explanatory variables. Unlike previous approaches that require a
hierarchical structure
(\protect\hyperlink{ref-carrizosa2022tree}{Carrizosa et al. 2022}) or
specific latent matrix factorizations
(\protect\hyperlink{ref-cerda2020encoding}{Cerda and Varoquaux 2020}),
our approach is amenable to the inclusion of group-level structured and
unstructured data, with the latter being an important novelty given the
prevalence and availability of high quality pre-trained embeddings. In
such cases, the method adopts the lens of first developing features in a
high-dimensional embedding and then rank reducing to balance accuracy
and precision.

We proceed as follows: The next section formalizes the econometric
considerations and illustrates core concerns through theoretical
analyses. We present simulation results from a running example,
illustrating that contemporary treatments rooted in well-accepted
practices may yield inaccurate inference. We introduce a novel program
of categorical embeddings in which each level of a categorical variable
is matched with a location in lower-dimensional coordinate system. We
demonstrate how features derived from this embedding may be useful in
establishing a novel econometric model enabling causal inference. We
compare analyses using real-world data in which we showcase empirical
findings from both existing models and practices and our proposed
estimation schema. The final section concludes.

\hypertarget{model-setup}{%
\section{Model Setup}\label{model-setup}}

As a key example, we consider the partially linear regression model as
described by Chernozhukov et al.
(\protect\hyperlink{ref-chernozhukov2018double}{2018}):

\[y_{i} = \alpha + f(x_i; \zeta) + \sum_{l=1}^{L_n} \beta_l D_{li} + \epsilon_i.\]

Here, \(i\) denotes an index variable, \(y_{i}\) represents the
dependent variable, \(\alpha\) is the intercept, \(f(x_i; \zeta)\)
signifies a smooth function of the observable variables \(x_i\)
parameterized by \(\zeta\), \(\beta_l\) represents the fixed
effect---the coefficient associated with the dummy variable \(D_{li}\),
and \(\epsilon_i\) is the error term. This model is typical in all
respects except for two aspects: (1) the superscript \(L_n\), indicating
that the number of levels captured by the dummy variables \(D_{li}\) is
a function of the sample size \(n\); and (2) we consider cases where
\(\beta\) is critical to the estimation challenge.

A sample-size-dependent \(L_n\) plays a role for the following reasons.
In cases where \(L_n\) remains constant, as classically expected,
\(\beta\) is identified in most instances, and the model is relatively
straightforward to estimate. However, in the scenario where \(L_n = n\),
with a novel category level generated for each observation, the model
becomes underidentified as only one observation corresponds to each
level. Consequently, \(\zeta\) and \(\beta\) present with sufficient
degrees of freedom to perfectly fit each observation.

When might such a structure emerge? Consider a merchant selling products
direct to consumers. As the business grows, so does the diversity of
addresses. In a categorical variable representing a consumer's zip code,
the number of included category levels would, for all practical
purposes, grow with the sample size as long as more data is generated
and added from consumers in different zip codes.

Or, in education sciences, consider assessments of educational efficacy
in school districts or teacher programs. A longitudinal study tracking
student outcomes across multiple cohorts may start with a limited number
of school districts but gradually expand to include more districts as
the study progresses. As more data is collected over time, the number of
distinct school districts or teacher education programs represented in
the data may increase
(\protect\hyperlink{ref-darling2005does}{Darling-Hammond et al. 2005},
\protect\hyperlink{ref-hanushek2016matters}{Hanushek 2016}). In a
categorical variable representing school districts, the inclusion of
these additional category levels, driven by the growing scope of the
data, would lead to the same challenges of high dimensionality and
sparsity.

Given there are only a finite number of zip codes or school districts in
the United States, it is conceivable that at some point, a categorical
variable of zip codes or school districts would saturate, and additional
data would only accrue for zip codes and school districts that were
already included in the model by virtue of earlier observations.
However, in most practical data samples, for instance those with fewer
than 41,700 observations in data where there are 41,700 zip codes and
all zip codes are candidates for inclusion, saturation will not occur.

Moreover, while the coding schema for zip codes and school districts is
well-defined and finite, the complexity of real-world data often
presents scenarios with no upper limit to the number of categorical
levels. For instance, consider including `brand' as a categorical
variable in our merchant example. As the product line evolves over time,
with new brands being introduced and older brands becoming dormant, the
number of levels in this variable (\(L_n\)) will inherently increase
with the sample size (\(n\)) and become potentially unbounded. This
phenomenon is not limited to brands alone; similar observations can be
made for store locations, product attributes, and other variables---all
of which may vary over time such that new values emerge and old values
become irrelevant. As such, any dynamic element in the research
environment that is described using a categorical variable will
inevitably lead to a dynamically varying definition of categorical
variable levels as new levels are constructed to track emergent aspects
and extant levels become dormant with environmental changes.

Behind these assertions lies the expectation that \(\beta\) is crucial
to the research question. To further develop this example, consider a
scenario where a firm seeks to determine where it should focus its
marketing efforts. Using zip code as a categorical variable could serve
as a natural proxy to understand variations in factors such as demand
across different zip codes. In such cases, accurately estimating
\(\beta\) while maintaining the granularity of the corresponding
categorical variable may be essential to the analysis. However, while
researchers may consider devising alternate schemas---such as merging
`neighboring' category levels like adjacent zip codes---such approaches
would inevitably lead to a loss in the specificity of findings; a dance
between specificity and estimation robustness is foundational to the
specification of a categorical schema in the econometric model.
Additional examples may include the Fama-French model, where the
firm-specific intercepts (designated as `alpha' in the canonical
treatment) may often be more critical than the coefficients on the four
factors.\footnote{A firm-specific alpha (i.e., \(\alpha_f\) in the
  Fama-French literature) is represented as \(\alpha + \beta_f\) in our
  specification, where \(\beta_f\) denotes the deviation of firm \(f\)'s
  coefficient from the grand mean.}

Our arguments differ from prior explanations of inconsistency caused by
the inclusion of fixed effects, most prominently in panel data models.
For example, in our case of consumer purchases, if the data were an
unbalanced panel with a varying number of observations for each consumer
(panel) and we sought to include a consumer (panel) fixed effect,
imprecision in the panel fixed effect might lead to inaccurate inference
of a structural parameter---a concern known as the incidental parameters
problem (\protect\hyperlink{ref-lancaster2000incidental}{Lancaster
2000}).

Such setups are distinct from ours in that panel-specific fixed effects
are (1) not focal and are considered nuisance (or incidental)
parameters, and (2) represent an aggregation of the influence of many
unobserved panel-level factors. For instance, a consumer fixed effect
may correspond to many unobserved consumer-level factors such as
demographics and psychographics, which may be both unobserved and
unethical to infer. For privacy reasons, the data may exclude the
consumer's precise address, making it unethical to establish such a
variable from proxies. Other examples include protected classes such as
age, ancestry, color, disability, ethnicity, gender, HIV/AIDS status,
military status, national origin, pregnancy, race, religion, and
sex---all variables that might influence the consumer's decisions but
which may be illegal and unethical for a firm to record and use. In
these cases, the categorical variable fixed effect speaks to a latent
manifold.

As panel fixed effects are not pivotal, a common route to estimation is
differencing sequential observations to remove the influence of the
fixed effect---an approach that yields consistent population-level
parameter estimates at the expense of being uninformative about the
fixed effects (\protect\hyperlink{ref-arellano1991some}{Arellano and
Bond 1991}). For instance, similar methodology has been used to estimate
the effects of psychosocial job stressors on mental health while
accounting for individual-specific factors such as some individuals
being more likely to report better mental health than others
(\protect\hyperlink{ref-milner2016persistent}{Milner et al. 2016}). This
solution, however, is less desirable when the manifold can be formed
through structured and unstructured variables such that the categories
are on or near it, as then the categorical variable fixed effects can be
recovered in addition to the population parameters.

A partially linear setup is typical in high-dimensional categorical
variables. More sophisticated variants, such as the inclusion of
interactions between \(f(x_i; \zeta)\) and \(D_{li}\), can be considered
in our framework and would likely yield model structures that also hew
close to established econometric models. However, the identification
challenge stems from the high dimensionality of
\(\{\beta_l \}_{l=1}^{L_n}\). Our approach innovates by providing a
higher-order model structure to reduce data demands and drive
identification. While such structures can be ported to more complex
cases, this would come at the cost of increased data demands that may be
simpler to list in theory than to satisfy in practice; depending on the
dimensionality of the categorical variable and the underlying manifold,
more complex models may or may not be practicable even with our
innovations.

\hypertarget{dudleys-entropy-integral}{%
\paragraph{Dudley's Entropy Integral}\label{dudleys-entropy-integral}}

Consider \(x_i\) as occurring within a second countable Hausdorff space.
This implies that the space of observables is separable (it contains a
countable dense subset whose closure encompasses the entire space),
regular (for any closed set and a point not within it, there exist
disjoint open sets containing each), and normal (for any two disjoint
closed sets, there exist disjoint open sets containing them).
Furthermore, \(y_{i}\), \(\zeta\), and \(\beta\) are almost surely
confined to a compact domain. Under these conditions, let \((T, d)\)
denote the pseudo-metric space of estimation functionals, with \(X_t\)
being a centered Gaussian process indexed by \(T\) and measured with the
pseudo-metric \(d\). Then, the following holds
(\protect\hyperlink{ref-vershynin2018high}{Vershynin 2018}):

\[\mathbb{E}\left[\sup_{t \in T} X_t\right] \leq C \int_0^{D} \sqrt{\log N(T, d, \varepsilon)} \,d\varepsilon,\]

where:

\begin{itemize}
\tightlist
\item
  \(\mathbb{E}\) denotes the expected value,
\item
  \(\sup\) represents the supremum (the least upper bound) of the
  process,
\item
  \(C\) is a universal constant,
\item
  \(N(T, d, \varepsilon)\) is the minimum number of balls of radius
  \(\varepsilon\) needed to cover \(T\) in the pseudo-metric \(d\),
\item
  \(D\) is a bound related to the diameter of the set \(T\) under the
  pseudo-metric \(d\).
\end{itemize}

Here, \(T\) represents the set of estimation functionals (i.e.,
functions of the empirical process), each corresponding to a unique
parameter in \(\beta\) and \(\zeta\). Given that \(\zeta\) has a fixed
cardinality, it follows:

\[\int_0^{D} \sqrt{\log N(T, d, \varepsilon)} \,d\varepsilon \propto L_n.\]

\hypertarget{donsker-conditions-and-convergence}{%
\paragraph{Donsker Conditions and
Convergence}\label{donsker-conditions-and-convergence}}

The Donsker conditions consist of a set of sufficient criteria used to
establish that a properly scaled empirical process converges in
distribution to a Gaussian process---specifically, that it belongs to
the Donsker class under natural law
(\protect\hyperlink{ref-dudley2010universal}{Dudley 2010}). A key
requirement is that the Dudley entropy integral must be bounded---a
condition straightforwardly met when \(L_n\) is constant, as in the
canonical scenario where the categorical variable has a few levels fixed
at the outset of data collection.

However, in scenarios like those outlined earlier, such as when the
model incorporates brand-specific fixed effects and the data reflects
the evolution of a market, or when the research question involves
measuring Fama-French `alpha' factors as firms emerge and evolve over
time, the entropy integral becomes unbounded, and parametric complexity
is not static over time. Instead, modeling complexity increases as the
data grows, a direct consequence of introducing novel values for the
categorical variable with the addition of more data. Consequently, the
estimation process no longer aligns with the Donsker conditions.

The Donsker conditions are sufficient but not necessary. However, the
intuition introduced by the Donsker conditions facilitates constructing
scenarios in which estimators are likely to converge and those in which
they are not.

For instance, consider a scenario where there is a single categorical
variable with a new level introduced every \(N > 1\) observation, and
each category level is equally likely to occur in every future
observation. In this scenario, which maps onto our simulations and
application, the category level fixed effect estimators are imprecise
but consistent. Alternatively, suppose a new level is introduced in each
observation. In this scenario, each observation corresponds to a
parameter, making the estimator of each fixed effect inconsistent (the
addition of more data does not reduce estimation error), although
estimators across fixed effects (and hence observations) may be
consistent if the errors in fixed effect estimates cancel out. For
example, a mean fixed effect, or a group-mean fixed effect if
considering hierarchical grouping, may be consistent even if the
underlying fixed effects are not. Thus, consistency when the Donsker
conditions are violated is nuanced.

This issue also presents a detection challenge because if the functional
central limit theorem does not apply, then the test statistics, which
are functionals of the empirical process, may fail to converge. For
example, in the pathological case where each observation corresponds to
a distinct fixed effect, the model will vastly overfit any data sample.
However, in contrast to typical applications where overfitting can be
mitigated by adding more data, in scenarios where new or additional data
does not correspond to extant fixed effect estimates, the model will
`overfit' in the asymptotic case.

Crucially, the R-squared test statistic becomes meaningless as the
second moment of the residuals is always 0 (due to the model being
overparametrized to the extent that it fits each observation perfectly),
and it never converges to the true variance of the structural error. As
such, even if one category level becomes dormant in the sense that for
some \(N^\ast\), observations \(n > N^\ast\) do not correspond to the
category level, the estimator of the variance of the structural errors
is asymptotically biased as the estimator of the dormant category level
is asymptotically biased---an issue that may be difficult to detect when
dealing with big data and categorical systems with thousands of
categories. In such cases, it becomes necessary to reformulate the
econometric model to ensure that parametric complexity does not depend
on the data---an approach we adopt in the model we propose.

\hypertarget{categorical-embedding}{%
\paragraph{Categorical Embedding}\label{categorical-embedding}}

Returning to our merchant example, consider embedding zip codes in a
2-dimensional space where the axes correspond to longitude and latitude.
This approach may provide valuable insights. For example, if a retailer
sells apparel suited to specific temperatures (e.g., T-shirts),
projecting the high-dimensional categorical variable onto a
2-dimensional space and including each corresponding location as data
features could offer a nuanced understanding of the economic problem.
The model could be easily expanded by incorporating additional features,
such as elevation.

Imagine, in addition to elevation, we have data on attributes like
rainfall. Suppose we augmented the geographic variables with a large
language model (LLM) embedding of the verbal description of each zip
code. These data features may prove useful as they encompass not only
the geographic characteristics of each zip code but also information on
economics, culture, etc. Similarly, in our education sciences example,
other data features such as average teacher salary, student-teacher
ratio, and per-pupil funding may be useful for modeling the fixed
effects associated with each school district or teacher education
program, leveraging auxiliary information to capture the underlying
characteristics of these entities and enable more accurate and efficient
inference (\protect\hyperlink{ref-kane2008does}{Kane et al. 2008},
\protect\hyperlink{ref-rivkin2005teachers}{Rivkin et al. 2005}).
Furthermore, in cases where unstructured descriptions of the category
levels (e.g., school district, teacher program) are available---such as
verbal descriptions of various school districts---this information can
be used to augment the available structured information to further
enhance estimation. This approach aligns with the use of school-level
characteristics to improve the precision of estimates in education
production function models, as demonstrated by Hanushek and Rivkin
(\protect\hyperlink{ref-hanushek1996aggregation}{1996}).

However, the addition of further data features brings up a significant
methodological challenge. While many data features may prove
informative, the addition of additional features increases the
dimensionality of the manifold and, therefore, the estimation challenge.
While in some applications the data may be rich enough to conduct
inference on a truly high-dimensional set of data features---for
instance, an LLM embedding may be as much as 3,072 dimensions, but is
fixed in size and does not change with sample size, and also pales in
comparison to the millions of data points that are feasible in modern
big-data applications---in many applications, we may also benefit from
dimensionality reduction.

Moreover, many data features may be irrelevant. For instance, LLM
embeddings are designed to capture complex narratives across a broad
spectrum of topics, the majority of which are unlikely to be relevant to
the analysis. In a model of apparel sales in the United States,
dimensions used to encode information on Greek literature or Roman
architecture in an LLM embedding are likely to be less relevant.
Therefore, reducing the dimensionality of the embedding might be useful.

We advocate the use of principal components analysis (PCA) for this
purpose. We recommend PCA as it focuses on a linear transformation of
the relative differences between category levels as expressed in the
original high-dimensionality embedding, in a lower-dimensionality space.
The directions (principal components) capturing the largest variation
are retained such that each additional direction is the line that best
fits the data while being orthogonal to all prior directions. Thus, the
directions constitute an orthonormal basis.

Applied to the fixed effects in the partially linear model, this process
`squeezes' out dimensions in which the category levels do not
systematically vary, as the objective in fixed effects is solely to
explain the variation in the influence of the category levels. On the
other hand, dimensions that explain variation in category level
distances feature in the coefficients on the lower-dimensional
embedding. To the extent the lower-dimensionality embedding provides a
perfect capture of the high-dimensional embedding, we obtain a
restatement of our original model with no information loss even as the
parametric space is substantially reduced through the representation
compression provided by PCA. To the extent the lower-dimensionality
embedding is an approximation, we obtain a near reconstruction.

Thus, in sum, we propose the development of a categorical embedding by
(1) ingesting descriptions of the category levels (e.g., the names and
locations of zip codes), (2) processing any unstructured elements
through an LLM encoder to generate an LLM embedding, and (3) applying
PCA to reduce the rank of the high-dimensional embedding. This data
pipeline is both relatively straightforward, which is likely to aid
adoption by applied researchers and industry, and has the conceptual
advantage of mapping points in the ambient categorical space to a
lower-dimensional space such that local distances on or near a smooth
nonlinear manifold are expressed accurately.

\hypertarget{estimated-model}{%
\paragraph{Estimated Model}\label{estimated-model}}

Utilizing the embedding of categorical variables, we estimate the model
as follows:

\[y_{i} = \alpha + f(x_i; \zeta) + \sum_{l=1}^{L_n} {\sum_{j=1}^{J} \gamma_j \hat{x}_{lj} D_{li}}  + \epsilon_i.\]

This formulation represents the expansion where
\(\beta_l = \sum_{j=1}^{J} \gamma_j \hat{x}_{lj}\), and
\(\gamma = \{\gamma_j \}_{j=1}^J\) are the coefficients corresponding to
the \(J\) dimensions of the rank-reduced embedding, representing the
categories. Here, \(\hat{x}_{l} = \{\hat{x}_{lj} \}_{j=1}^J\) specifies
the location of the \(l^{\text{th}}\) categorical level within the
embedding.

This model formulation, by design, is bounded in complexity due to the
coefficient vector's limited cardinality. Furthermore, all categorical
variables are projected onto the \(J\) dimensions, ensuring
dimension-related information is accumulated across all observations for
all categorical levels, thus addressing the issue of categorical level
sparsity. Asymptotic normality and a convergence of the estimation
functionals follow as the model is structurally identical to a classical
regression. After estimation, recovering estimates and standard errors
in the original model is straightforward, based on the estimates and
distributions of \(\gamma_j\), and the embedding of the categorical
variable.

It is straightforward to extend this model to the case
\(\beta_l = h ( \hat{x}_{l} ; \gamma )\) where \(h\) is a smooth
function and \(\gamma\) continue to be coefficients. A reason to be
cautious, however, is that the dimensionality of the embedding \(J\) is
chosen to balance the data demands of the model with precision. It is in
this sense that the specification of \(h\), whether a linear function as
we advocate or a more flexible function, plays an additional role as
flexibility in \(h\) may entail a loss of detail in the embedding
through smaller \(J\). Ultimately, though, these are relatively
straightforward issues that can be resolved in a research question and
context through empirical testing.

\hypertarget{simulations}{%
\section{Simulations}\label{simulations}}

We conduct the following simulations within the framework outlined in
the previous section. In our first simulation, we explore a scenario
where a single covariate represents the net price offered to a consumer
(after accounting for discounts and coupons), and a single categorical
variable denotes the zip code. The dependent variable is the total
revenue generated from a single consumer located in the specified zip
code within a given period. We assume a coefficient of -1 on price to
reflect the conventional understanding that the demand curve is downward
sloping.

The category in our simulations is summer apparel, which matches the
category in our empirical analysis on real-world data reported in the
next section. The coefficient on the categorical variable (zip code) is
modeled as a linear function of the latitude and longitude of a zip
code's centroid. We assume changes in longitude have no effect on
revenues, mirroring a scenario focused on summer apparel, which is not
systematically affected by location east-west. Conversely, we assign a
coefficient of -1 to latitude, indicating decreased demand in the north
due to colder climates. The error term is assumed to follow a standard
normal distribution, with one observation per consumer (i.e., the data
is cross-sectional). Lastly, we standardize all geographic variables by
subtracting the minimum and dividing by the standard deviation, and set
the intercept to 0 to simplify the exposition.

This setup is particularly representative of direct marketing, where
personalized and targeted pricing as enabled through discounts and
coupons, occurs by zip code (i.e., prices net of coupons are set by zip
code, e.g.~see \protect\hyperlink{ref-frana2023demographics}{Frana
2023}). To capture its stochastic nature, we draw price from a uniform
distribution between 0 and 10. We add to this draw the difference
between the maximum latitude in our data and thus the northernmost zip
code in the United States and the latitude of a zip code. This is meant
to capture price's systematic nature where price is often correlated
with a variable that is excluded in the main model but that may enter
through the fixed effect (in this case, latitude). Thus, we simulate
prices as likely to be higher in Florida than in Maine, reflecting a
higher demand (and therefore higher price) for summer apparel---i.e., in
our simulation, we assume that consumers in more northern zip codes are
offered systematically lower prices through couponing and price
personalization. The converse assumption, however, is also tenable where
increased demand may lead to systematically lower prices; for the
purpose of investigating our methodology, the key element is simulating
price endogeneity, which is achieved by our design choice. The
alternative is straightforward to study using our model and should lead
to substantively similar findings on the impact of various estimators on
estimate accuracy and precision.

We randomly sample zip codes based on the population in each zip code.
This process results in a categorical variable that is both dynamically
increasing and sparse within our dataset of 100,000 simulated
observations, due to the data being insufficient to fully saturate the
categorical data space. We employ data as obtained and analyzed using
open-source tools for the simulation to enhance replicability. To that
end, we use the package `zipcodeR' in R. To simplify the analysis, given
the total number of zip codes is too large for analysis using our
simulation whereby sampling from all zip codes would yield an even more
sparse categorical variable, we drop zip codes for which there either is
no corresponding latitude and longitude because the zip code is for a
post box, and zip codes for which such variables are missing. We then
keep only the top 10,000 zip codes by population in the simulation. We
sample these zip codes by weight of population such that more populous
zip codes are sampled multiple times.

\begin{figure}[htbp]
\centering
\includegraphics[width=0.5\linewidth]{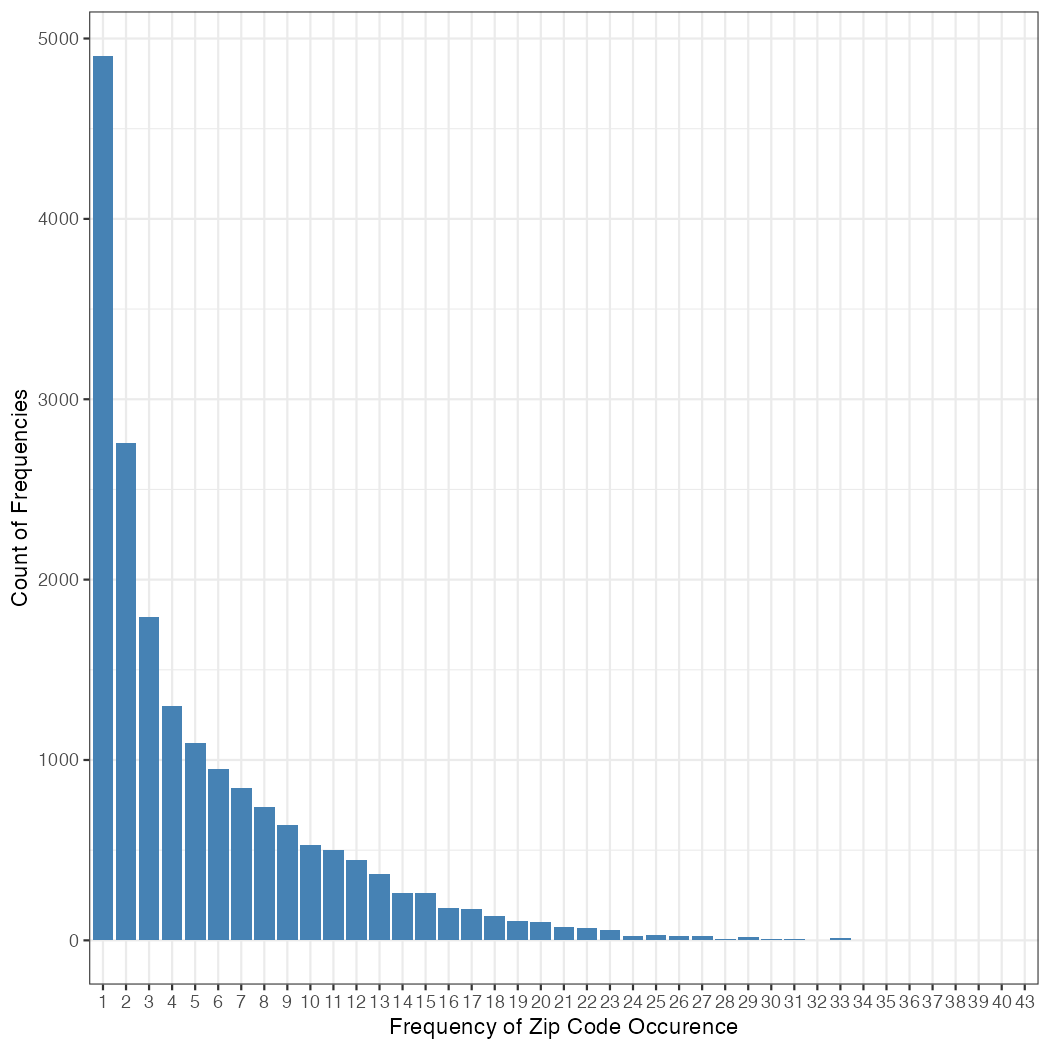}
\caption{Distribution of Zip Code Frequencies}
\label{fig:frequency}
\begin{minipage}{\linewidth}
\medskip
\footnotesize
Note: The x-axis represents the frequency with which a zip code occurred in our simulation data. The y-axis shows the count of zip codes that occurred with each given frequency.
\end{minipage}
\end{figure}

Figure \ref{fig:frequency} illustrates the frequency table of the zip
codes in our data. On the x-axis is the frequency with which a zip code
appears in the data, and on the y-axis is the count of zip codes that
occurred with a given frequency. For instance, among the least popular
zip codes, 5,031 zip codes were observed once, and 2,576 zip codes were
observed twice. Among the most popular zip codes, 2 zip codes were
observed 38 times each, 1 zip code was observed 40 times, and 1 zip code
was observed 46 times. In total, 18,468 zip codes were selected at least
once. Thus, the data describe a power law in the frequency with which
zip codes occur, as expected, given that the basis on which the zip
codes were sampled---the population distribution across zip codes---also
follows a power law. We simulate this data on the basis of a marketing
research context and question but similar datasets are likely to emerge
in a host of social science settings. For instance, in any data that is
randomly sampled from the U.S. population and then binned to the zip
code to provide anonymity, we would expect the emergence of zip codes in
the data to follow the same process as in our simulation.

Our econometric model is formulated as follows:

\[y_i = \alpha + \beta_p x_{ip} + \sum_z \beta_z D_{iz} + \epsilon_i,\]

where:

\begin{itemize}
\tightlist
\item
  \(\alpha = 0\), \(\beta_p\), and \(\{\beta_z\}_z\) represent the
  intercept, the coefficient on the price index, and the coefficients on
  the categorical variable (i.e., zip code fixed effects), respectively.
\item
  \(x_{ip}\) denotes the price.
\item
  \(D_{iz}\) is an indicator variable that equals one if consumer \(i\)
  is located in zip code \(z\); for simplicity, we use \(\sum_z\) to
  denote a sum over all zip codes included in the model.
\item
  \(\epsilon_i\) is the error term, assumed to follow a standard normal
  distribution.
\end{itemize}

In this specification, zip codes are treated according to their observed
ambient space. Therefore, in data with 18,468 zip codes, it necessitates
the inclusion of 18,468 category levels, resulting in a specification
with 18,468 fixed effects. However, this space is underpinned by a
manifold of latitude and longitude that determines the consumer's
location and, consequently, \(D_{iz}\) (i.e., the zip code). Thus, we
further specify
\(\beta_z = \delta_{\text{long}} i_{\text{long}} + \delta_{\text{lat}} i_{\text{lat}}\),
where \(\delta_{\text{long}}\) and \(\delta_{\text{lat}}\) are the
coefficients on longitude and latitude on the manifold, respectively,
and \(i_{\text{long}}\) and \(i_{\text{lat}}\) represent the consumer's
longitude and latitude, respectively.

Our second simulation introduces elevation as a further dimension of the
manifold, while adhering to earlier simulation specifications. We obtain
the elevation of each zip code by extracting point elevations from the
`AWS Terrain Tiles', using the library `elevatr' in R. As each latitude
and longitude automatically defines elevation, we employ the same
locations as in the first simulation but with the fixed effect now
specified as
\(\beta_z = \delta_{\text{long}} i_{\text{long}} + \delta_{\text{lat}} i_{\text{lat}} + \delta_{\text{ele}} i_{\text{ele}}\),
and \(\delta_{\text{lat}} = \delta_{\text{ele}} = -1\), where
\(i_{\text{ele}}\) is consumer \(i\)'s elevation.

The key difference between the simulations is in the distribution of the
fixed effects. In the first simulation, the fixed effect ranges from 0
in the northernmost zip codes to 4.895 in Key West, FL. The mean fixed
effect is 2.305, and the standard deviation is 1. In the second
simulation, the fixed effect ranges from 2.025 (in Buena Vista, CO) to
15.839 (in Oxnard, CA). The mean is 12.054, and the standard deviation
is 1.518.

\hypertarget{results}{%
\subsection{Results}\label{results}}

\begin{figure}[htbp]
\centering
\begin{subfigure}{.5\textwidth}
  \centering
  \includegraphics[width=\linewidth]{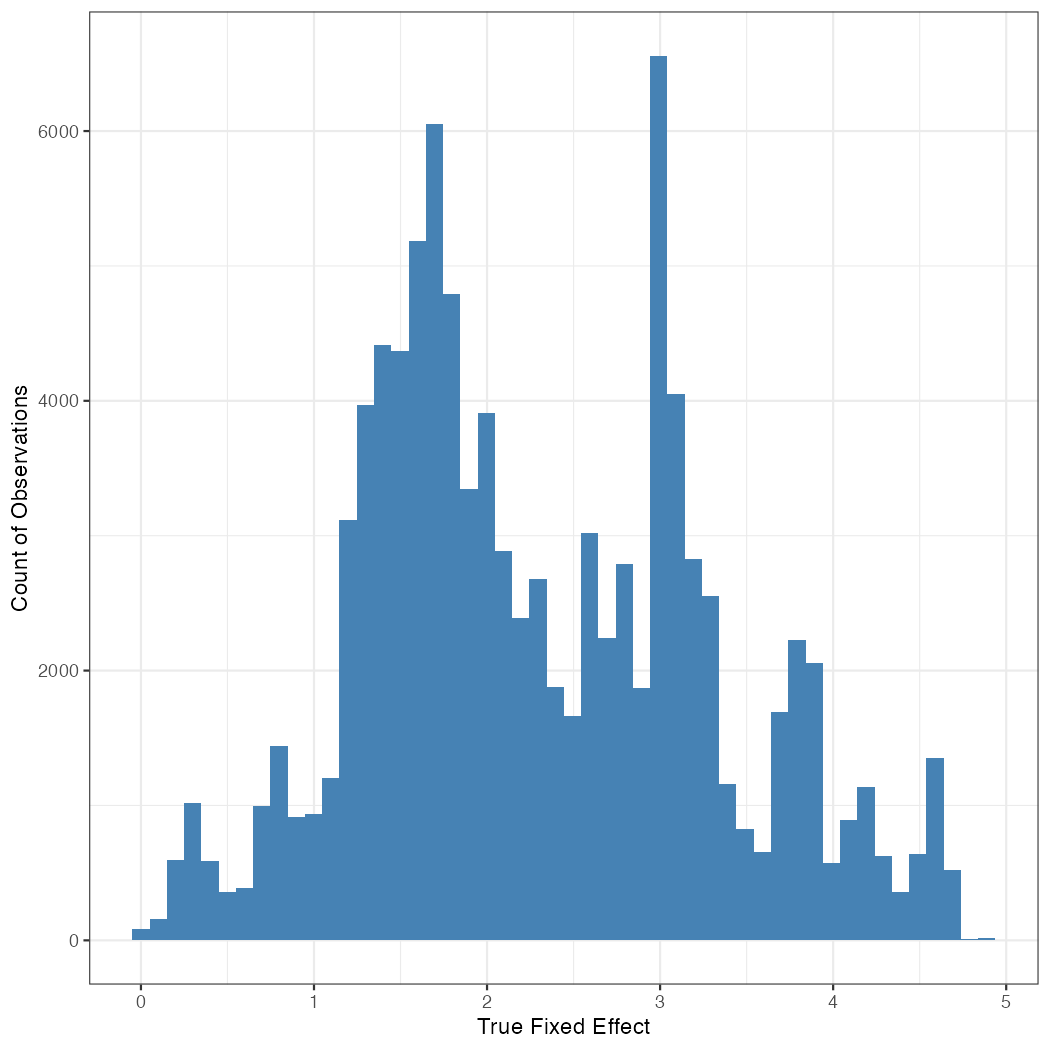}
  \caption{True Fixed Effects}
  \label{fig:estimate_against_true_1a}
\end{subfigure}%
\begin{subfigure}{.5\textwidth}
  \centering
  \includegraphics[width=\linewidth]{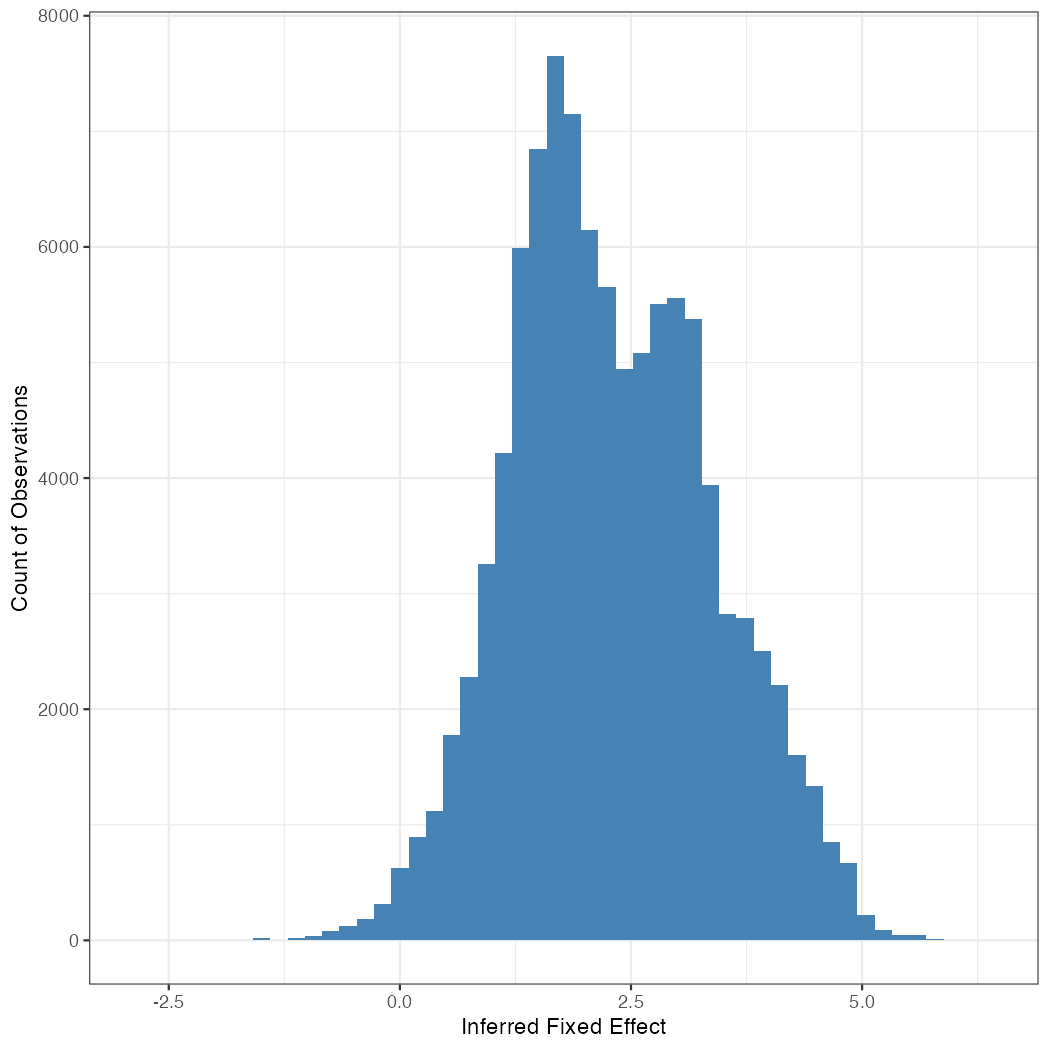}
  \caption{Regression: Inferred Fixed Effects}
  \label{fig:estimate_against_true_1b}
\end{subfigure}
\begin{subfigure}{.5\textwidth}
  \centering
  \includegraphics[width=\linewidth]{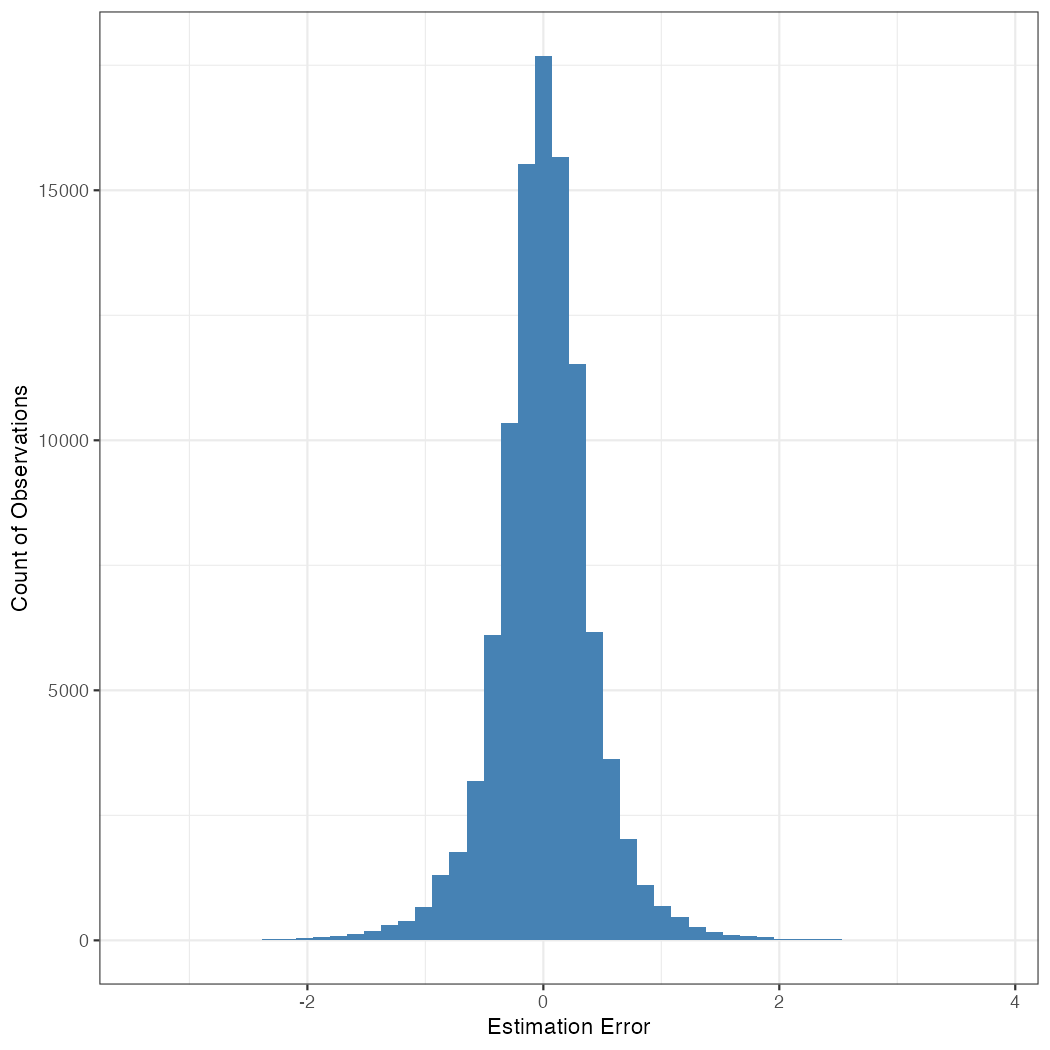}
  \caption{Regression: Estimation Error}
  \label{fig:estimate_against_true_1c}
\end{subfigure}%
\begin{subfigure}{.5\textwidth}
  \centering
  \includegraphics[width=\linewidth]{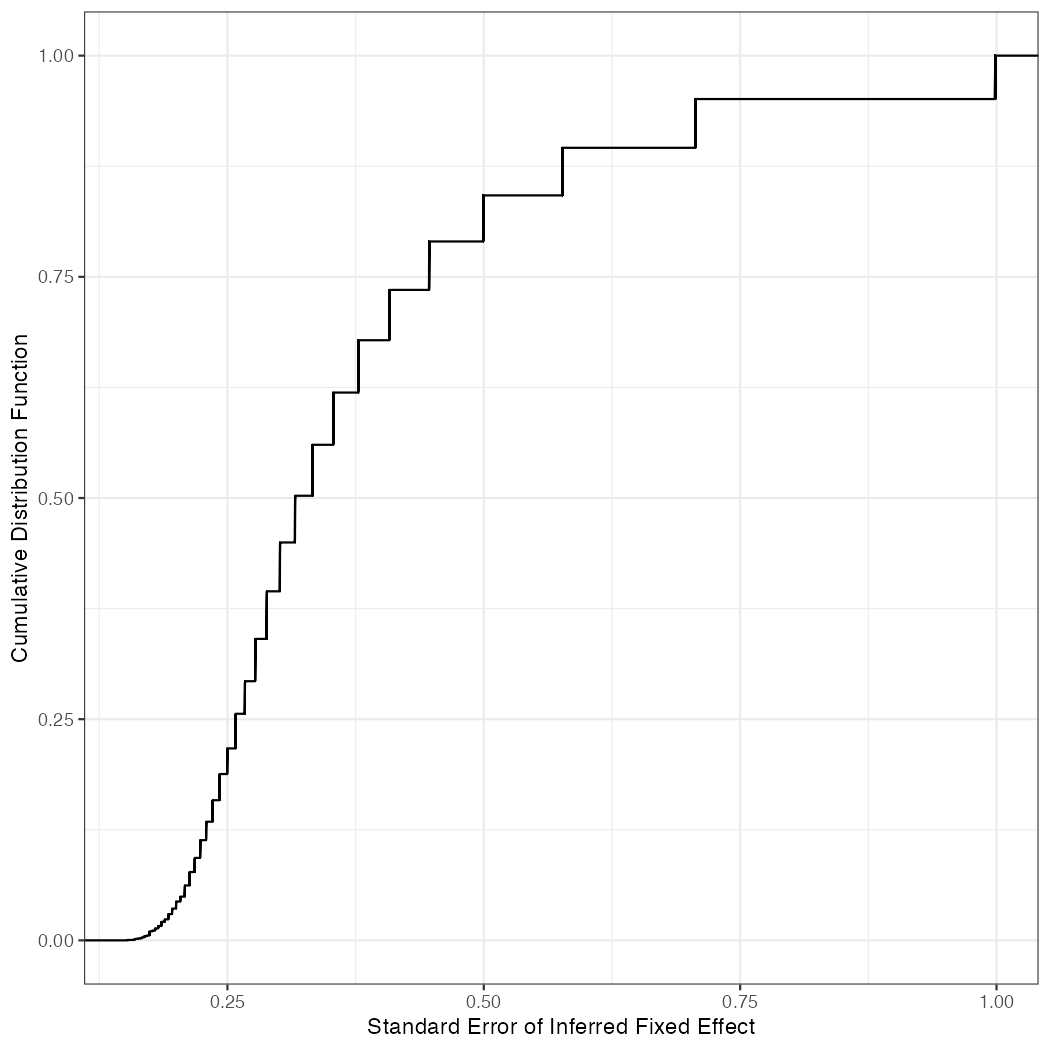}
  \caption{Regression: ECDF of Estimation Precision}
  \label{fig:estimate_against_true_1d}
\end{subfigure}
\caption{Regression Estimation Errors in Simulation 1}
\label{fig:estimate_against_true_1}
\begin{minipage}{\linewidth}
\medskip
\footnotesize
Note: Estimation accuracy and precision in simulation 1 for the fixed effect coefficients.
\end{minipage}
\end{figure}

\begin{figure}[htbp]
\centering
\begin{subfigure}{.5\textwidth}
  \centering
  \includegraphics[width=\linewidth]{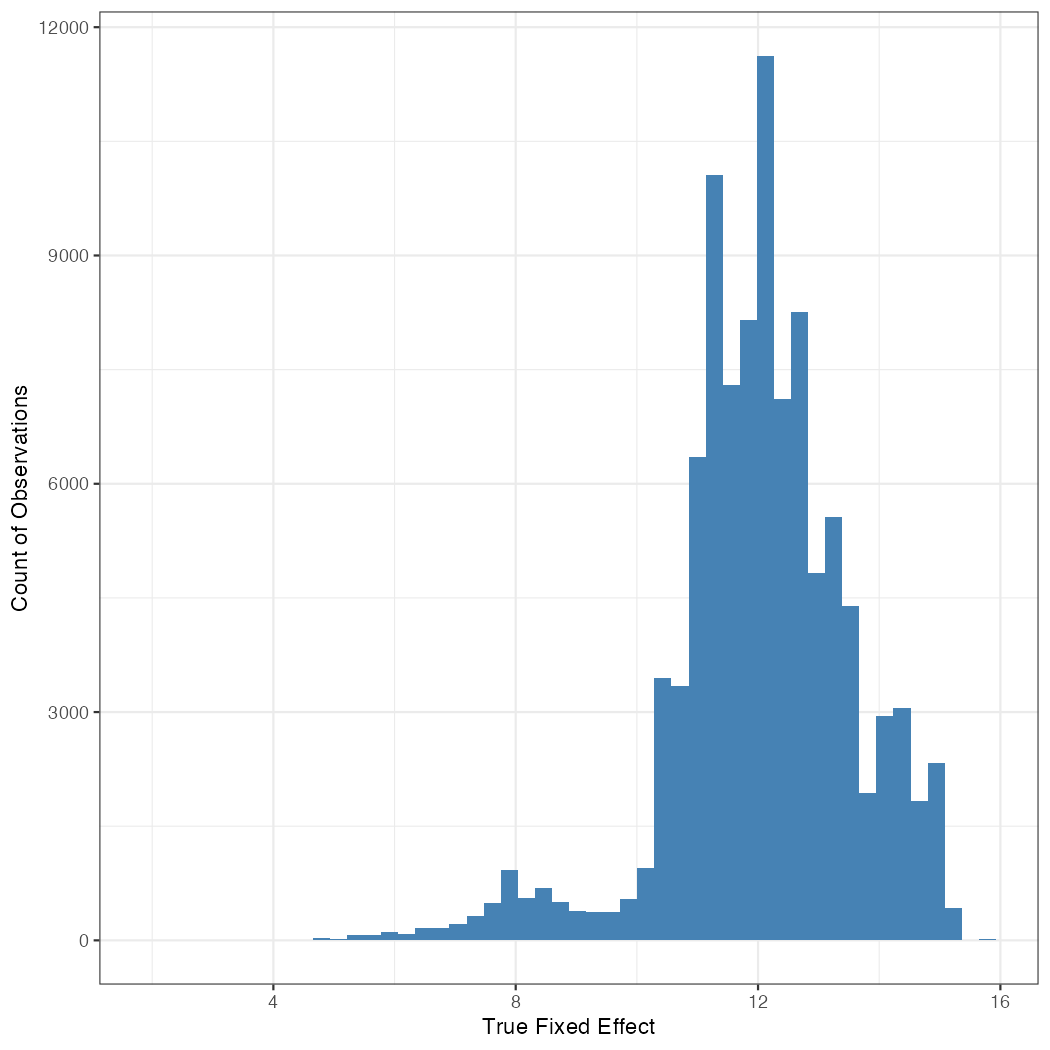}
  \caption{True Fixed Effects}
  \label{fig:estimate_against_true_2a}
\end{subfigure}%
\begin{subfigure}{.5\textwidth}
  \centering
  \includegraphics[width=\linewidth]{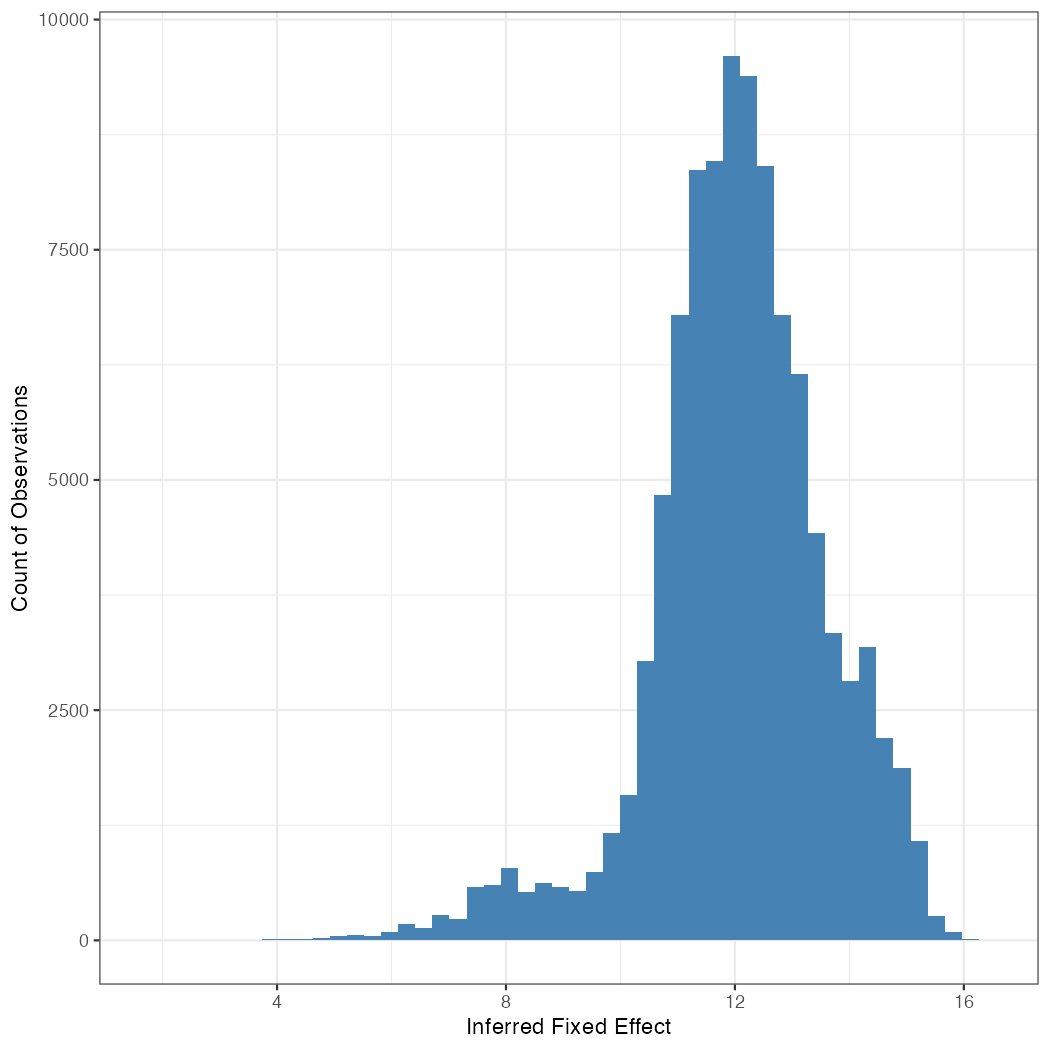}
  \caption{Regression: Inferred Fixed Effects}
  \label{fig:estimate_against_true_2b}
\end{subfigure}
\begin{subfigure}{.5\textwidth}
  \centering
  \includegraphics[width=\linewidth]{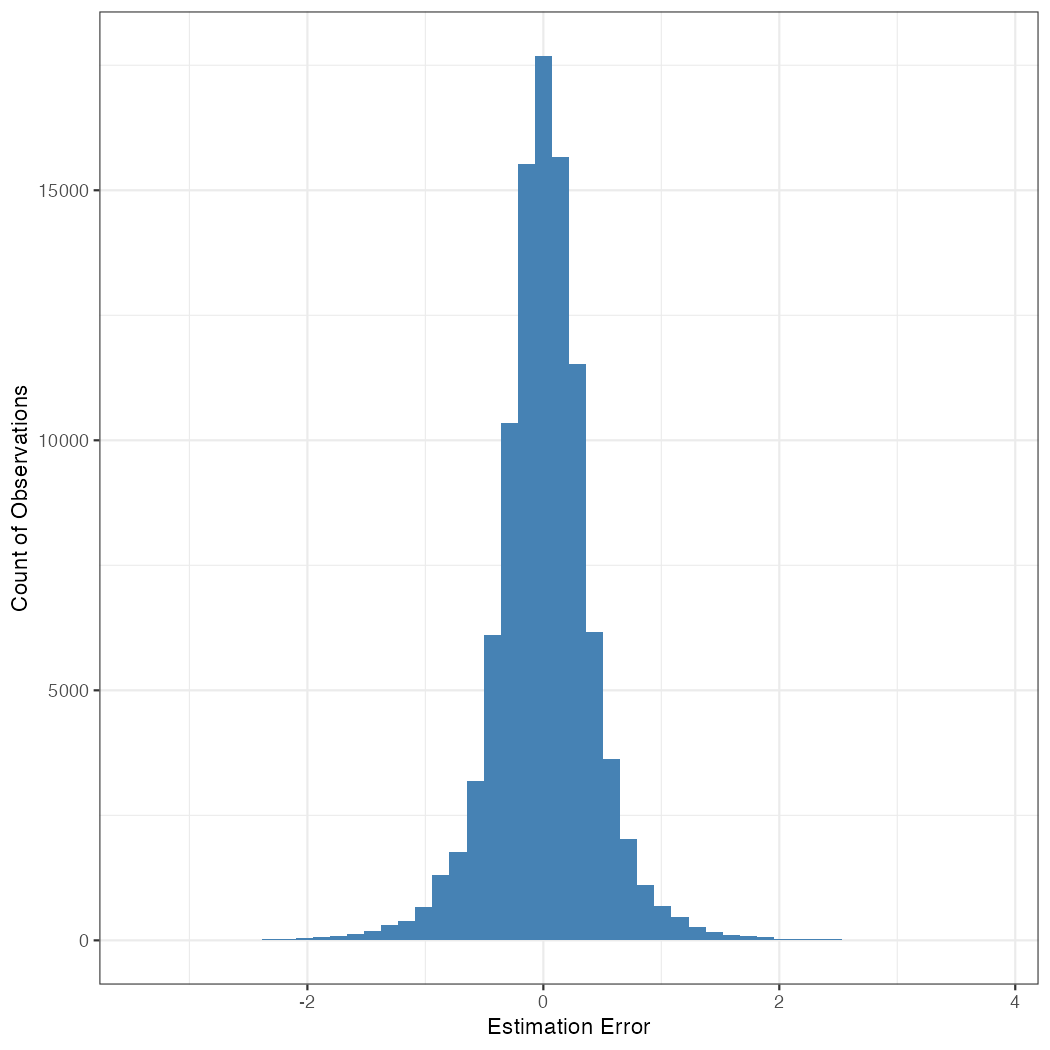}
  \caption{Regression: Estimation Error}
  \label{fig:estimate_against_true_2c}
\end{subfigure}%
\begin{subfigure}{.5\textwidth}
  \centering
  \includegraphics[width=\linewidth]{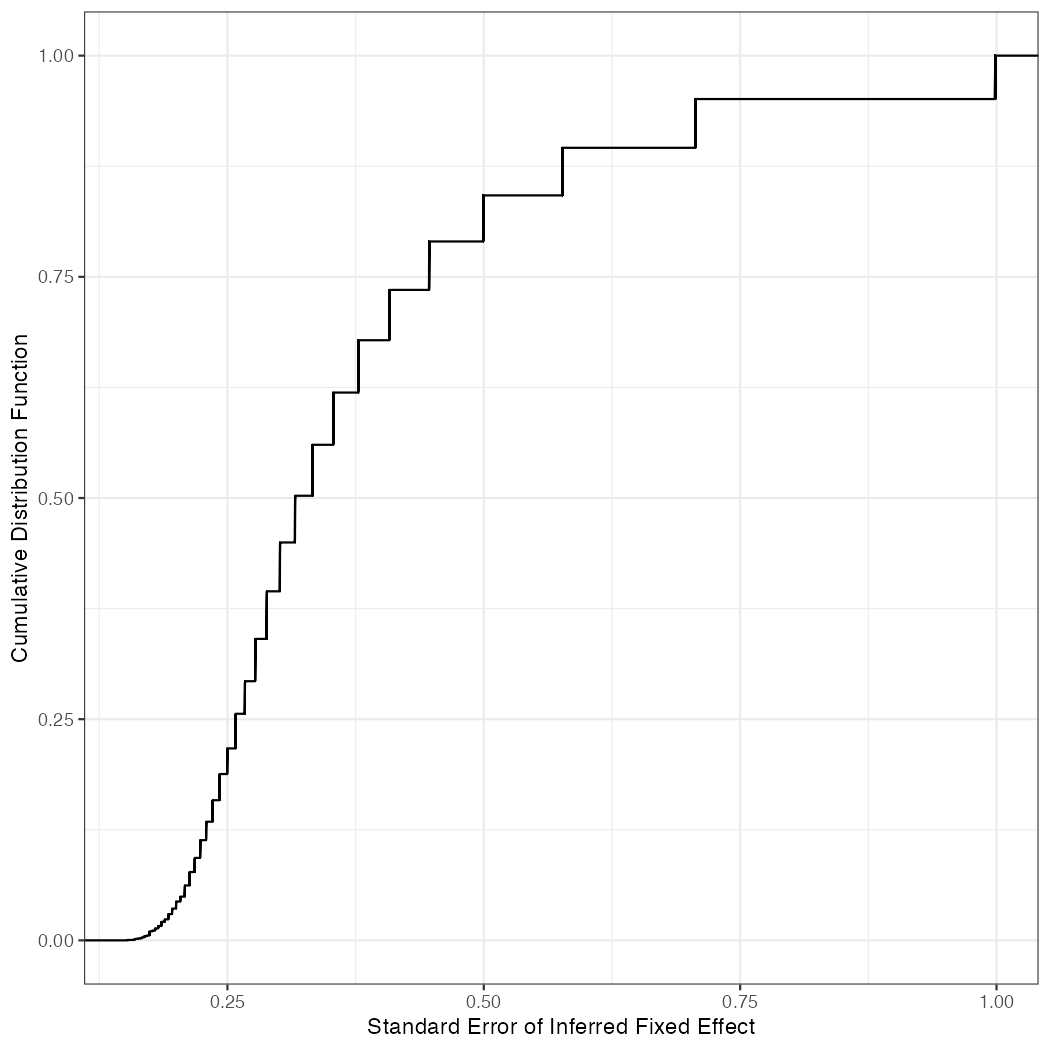}
  \caption{Regression: ECDF of Precision}
  \label{fig:estimate_against_true_2d}
\end{subfigure}
\caption{Regression Estimation Errors in Simulation 2}
\label{fig:estimate_against_true_2}
\begin{minipage}{\linewidth}
\medskip
\footnotesize
Note: Estimation accuracy and precision in simulation 2 for the fixed effect coefficients. ECDF = Empirical cumulative distribution function.
\end{minipage}
\end{figure}

Figure \ref{fig:estimate_against_true_1} and Figure
\ref{fig:estimate_against_true_2} characterize estimation accuracy and
precision in Simulations 1 and 2, respectively. They detail the true and
inferred fixed effect coefficients using the standard regression
estimator, as implemented by the `lm' function in R, which automatically
excludes category levels if the data are insufficient to provide a
non-singular fit. In all simulations, the coefficient on price is
estimated with great precision, as is to be expected in a simulation
with 100,000 observations and where price is drawn from a uniform
distribution.

Central to our research, and germane to our modeling advances, the
coefficient on price is non-focal in our simulation and assumes the role
of a nuisance variable. Instead, the analysis focuses on the
coefficients on the zip codes where the estimation challenge is the high
dimensionality of the categorical variable. In other cases, both
population parameters and category-level (or group-specific) parameters
may be crucial. We chose a simulation setup geared to illustrate the key
benefits of our proposed method, but our model and innovations are
applicable more broadly.

Subfigures \ref{fig:estimate_against_true_1a} and
\ref{fig:estimate_against_true_2a} present a density of the true fixed
effects, while subfigures \ref{fig:estimate_against_true_1b} and
\ref{fig:estimate_against_true_2b} present a density of the inferred
fixed effects. A visual comparison between these sets of subfigures
reveals that the density of the inferred fixed effects does not resemble
the density of the true fixed effects, despite the large sample size and
relatively simple econometric specification. The difference between the
density plots visually illustrates the issues we highlight in our
analysis, whereby many estimates pertain to category levels that are
estimated using only one or a few data points, leading to a distribution
of fixed effect estimates that is vastly distinct from the original true
fixed effects.

Subfigures \ref{fig:estimate_against_true_1c} and
\ref{fig:estimate_against_true_2c} present histograms of the estimation
error, which is the difference between the true fixed effect and the
inferred fixed effect. While each coefficient estimate is imprecise, the
expectation of the estimation error, taken over all fixed effects and
observations, is 0; thus, the estimates are, on net, consistent. This is
because in the multivariate regression, as long as there is some
non-zero probability of a category level re-expressing, the fixed effect
estimator is consistent. In our simulation, category levels (zip codes)
are not dormant. Therefore, the estimates are consistent, and the sample
average of estimation errors across coefficients is close to 0.

The distribution of the estimation errors is identical (to rounding
error) in subfigures \ref{fig:estimate_against_true_1c} and
\ref{fig:estimate_against_true_2c}. This is because
\(\hat{\beta_z} = \frac{\sum_i { D_{iz} (y_i - \hat{\beta_p} x_{ip}) }} {\sum_i {D_{iz}}}\).
Suppose \(y_{2i} = y_{1i} + \sum_z \delta_z D_{iz}\) for some
\(\delta_z\), with \(y_{2i}\) and \(y_{1i}\) being the \(i\)th
observation of the dependent variable in the second and first
simulations, respectively. It follows that
\(\hat{\beta_{2z}} = \hat{\beta_{1z}} + \delta_z\), where
\(\hat{\beta}\) is the regression estimator and \(\hat{\beta_p}\) is
estimated precisely. Thus, \(\delta_z\) only shifts the estimates---it
has no effect on estimator precision.

This feature of the fixed effects model illustrates a key reason for our
advocacy of PCA for dimensionality reduction. Namely, if the
lower-dimensional embedding is a linear projection of the ambient space,
as is the case with PCA, then the influence of each dimension of the
manifold persists through the linear projection and a linear estimator
such as the regression estimator. Consequently, each included or
excluded dimension in the lower-dimensionality embedding has the clear
and unambiguous meaning of contributing a projection of the fixed effect
to its estimation. The inclusion of superfluous dimensions leads to more
estimated null effects, greater data demands, and reduced precision
while preserving consistency, whereas the exclusion of critical
dimensions leads to biased estimates. For instance, projecting the fixed
effects in simulation 2 on the manifold in simulation 1 would yield
biased estimates due to the missing dimension (elevation).

Subfigures \ref{fig:estimate_against_true_1d} and
\ref{fig:estimate_against_true_2d} present the empirical cumulative
distribution function (ECDF) of the precision (i.e., the standard
errors) of the estimators. Here, we find variation in the extent of
precision, with the visual analog of a step function where each step
corresponds to a category level that found more expression in the data.
Mirroring the symmetry of subfigures \ref{fig:estimate_against_true_1c}
and \ref{fig:estimate_against_true_2c}, subfigures
\ref{fig:estimate_against_true_1d} and
\ref{fig:estimate_against_true_2d} are also indistinguishable.

If all estimators converge, then the mean of the square of the standard
errors of the fixed effects should converge to the variance of the
inferred fixed effects \(\hat{\beta}_z\) around their true mean
\(\beta_z\). The observed variance of the inferred fixed effects is
0.179 in our sample, whereas the expected precision (i.e., the average
of the squared standard errors) is 0.184. This deviation is
statistically significant as the standard error of the mean is 0.00216,
yielding a t-statistic of 2.23.

This distinction occurs because the model overfits in-sample. For
instance, the zipcode `01010' is expressed only once in the data. Its
inferred fixed effect is computed as \(y_i - \hat{\beta}_p x_{ip}\)
where \(i=48558\) is the corresponding observation; the fixed effect is
estimated using exactly one observation. It follows then that the
estimator is unable to ascertain its own precision as the residual is 0.
This implication of extreme sparsity is typical in power-law distributed
category levels.

\begin{figure}[htbp]
\centering
\begin{subfigure}{.5\textwidth}
  \centering
  \includegraphics[width=\linewidth]{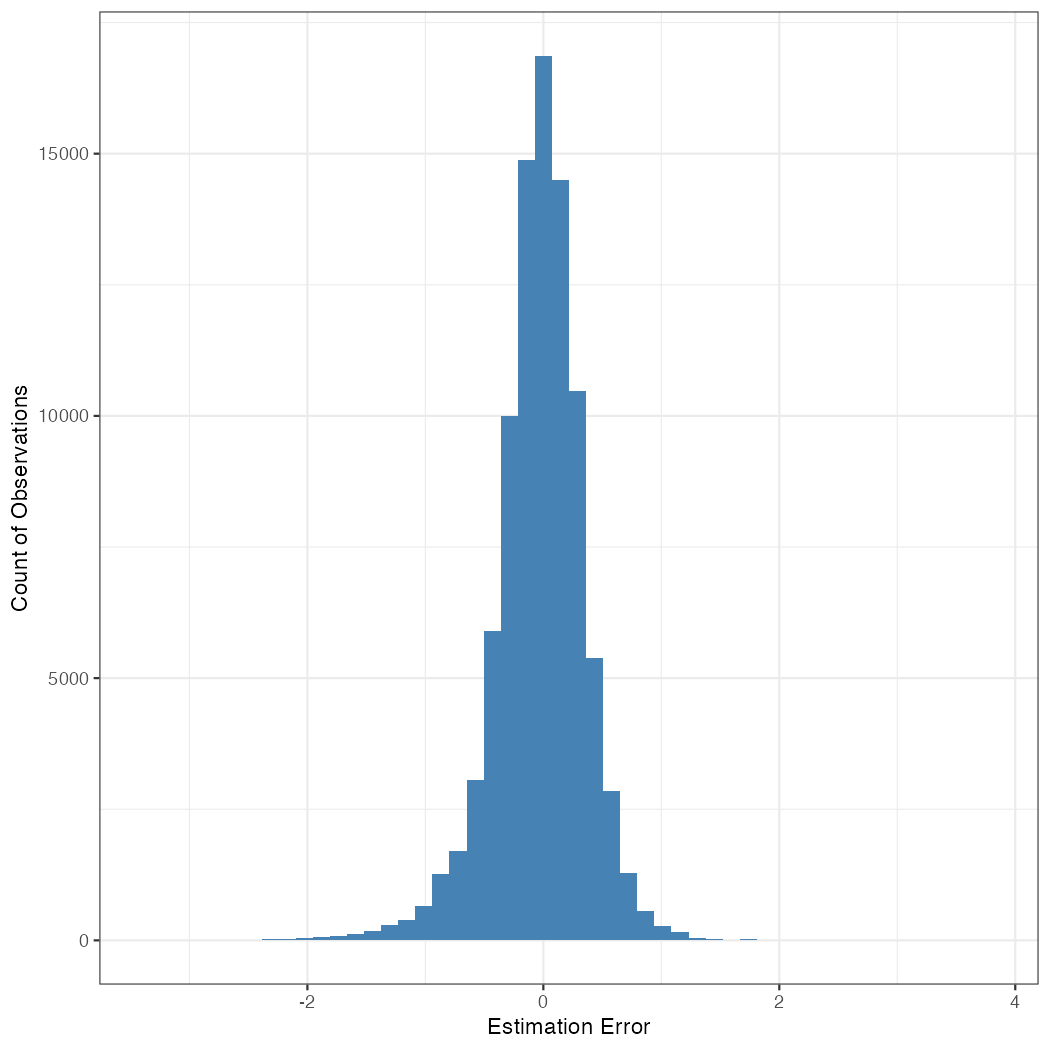}
  \caption{Significant Fixed Effects}
  \label{fig:estimate_against_true_3a}
\end{subfigure}%
\begin{subfigure}{.5\textwidth}
  \centering
  \includegraphics[width=\linewidth]{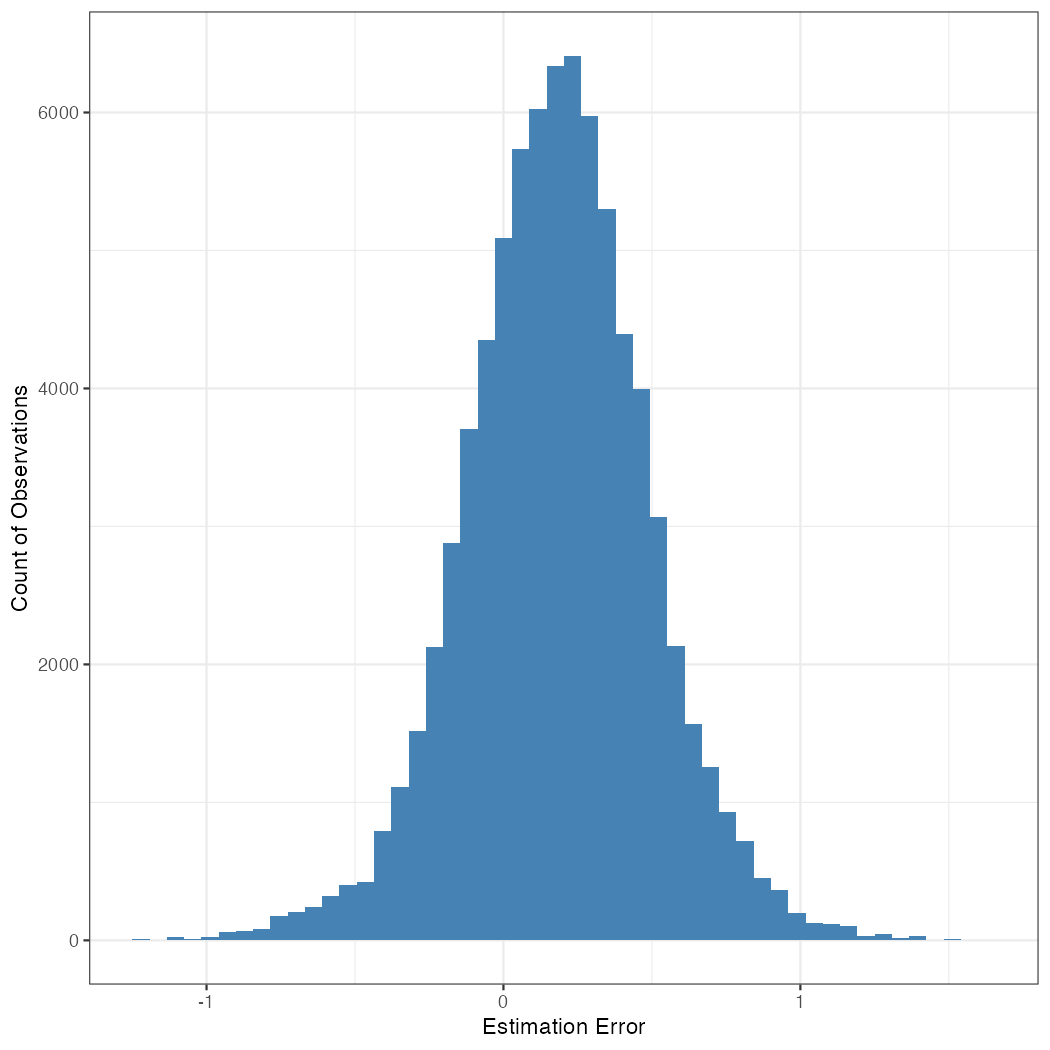}
  \caption{Merged Category Levels}
  \label{fig:estimate_against_true_3b}
\end{subfigure}
\begin{subfigure}{.5\textwidth}
  \centering
  \includegraphics[width=\linewidth]{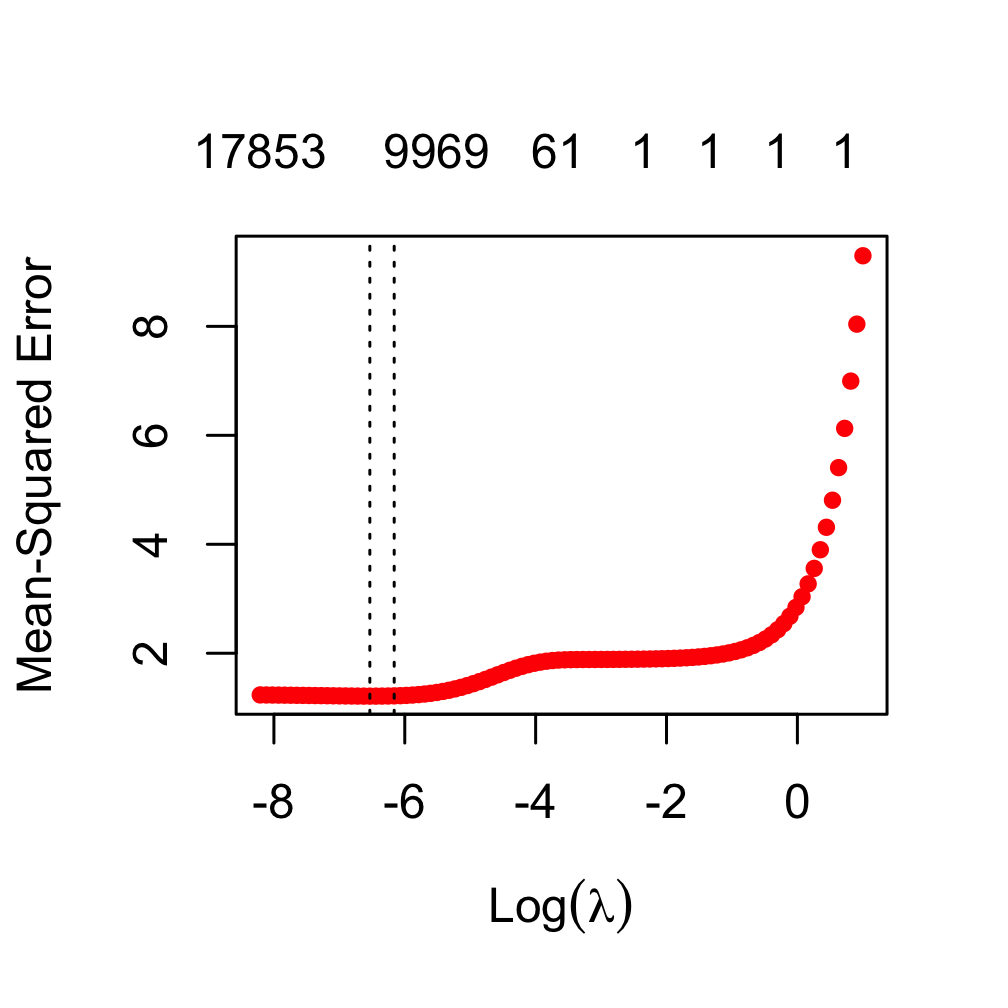}
  \caption{Lambda Selection}
  \label{fig:estimate_against_true_3c}
\end{subfigure}%
\begin{subfigure}{.5\textwidth}
  \centering
  \includegraphics[width=\linewidth]{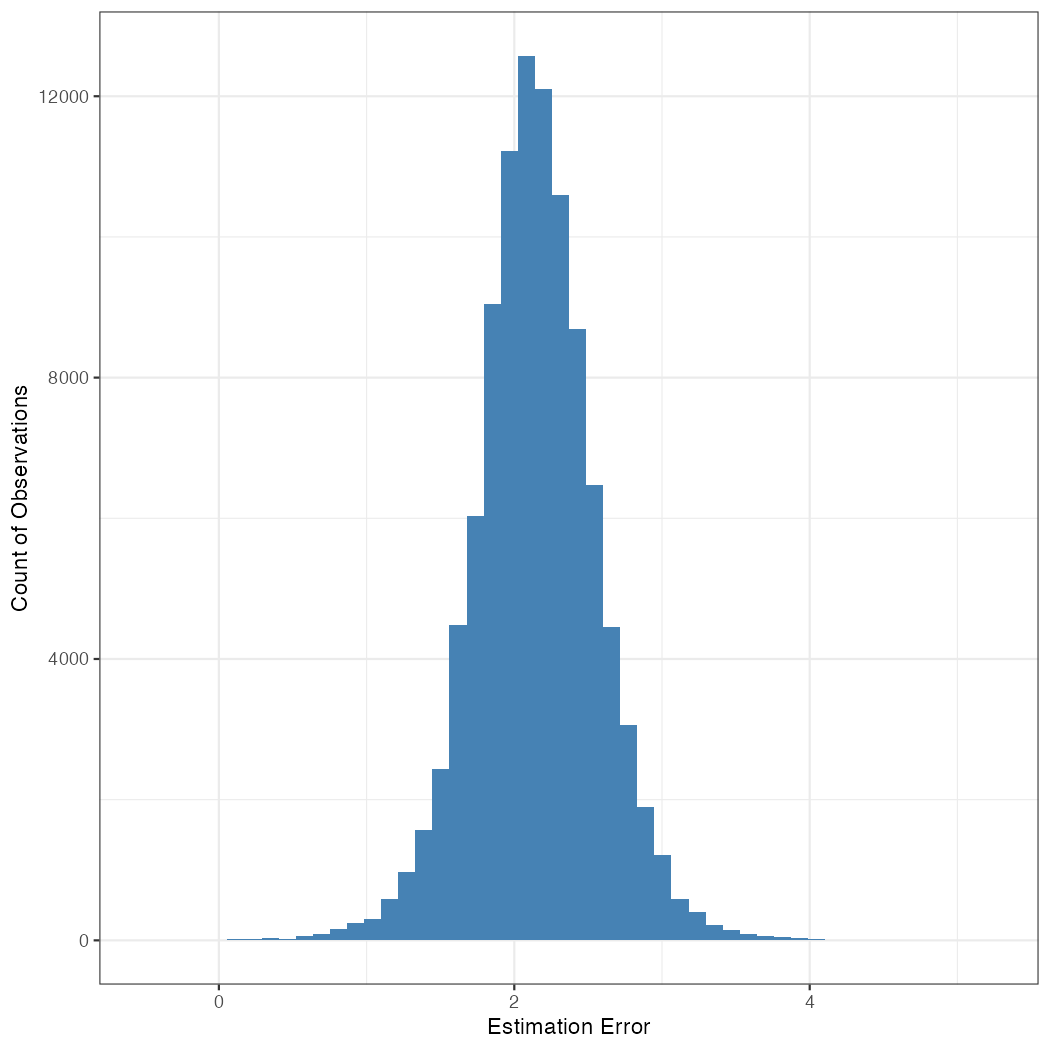}
  \caption{Estimation Errors, LASSO}
  \label{fig:estimate_against_true_3d}
\end{subfigure}
\caption{Simulation 1: Estimation Errors in Regression Analysis}
\label{fig:estimate_against_true_3}
\begin{minipage}{\linewidth}
\medskip
\footnotesize
Note: Estimation accuracy and precision in simulation 1 for the fixed effect coefficients.
\end{minipage}
\end{figure}

\begin{figure}[htbp]
\centering
\begin{subfigure}{.5\textwidth}
  \centering
  \includegraphics[width=\linewidth]{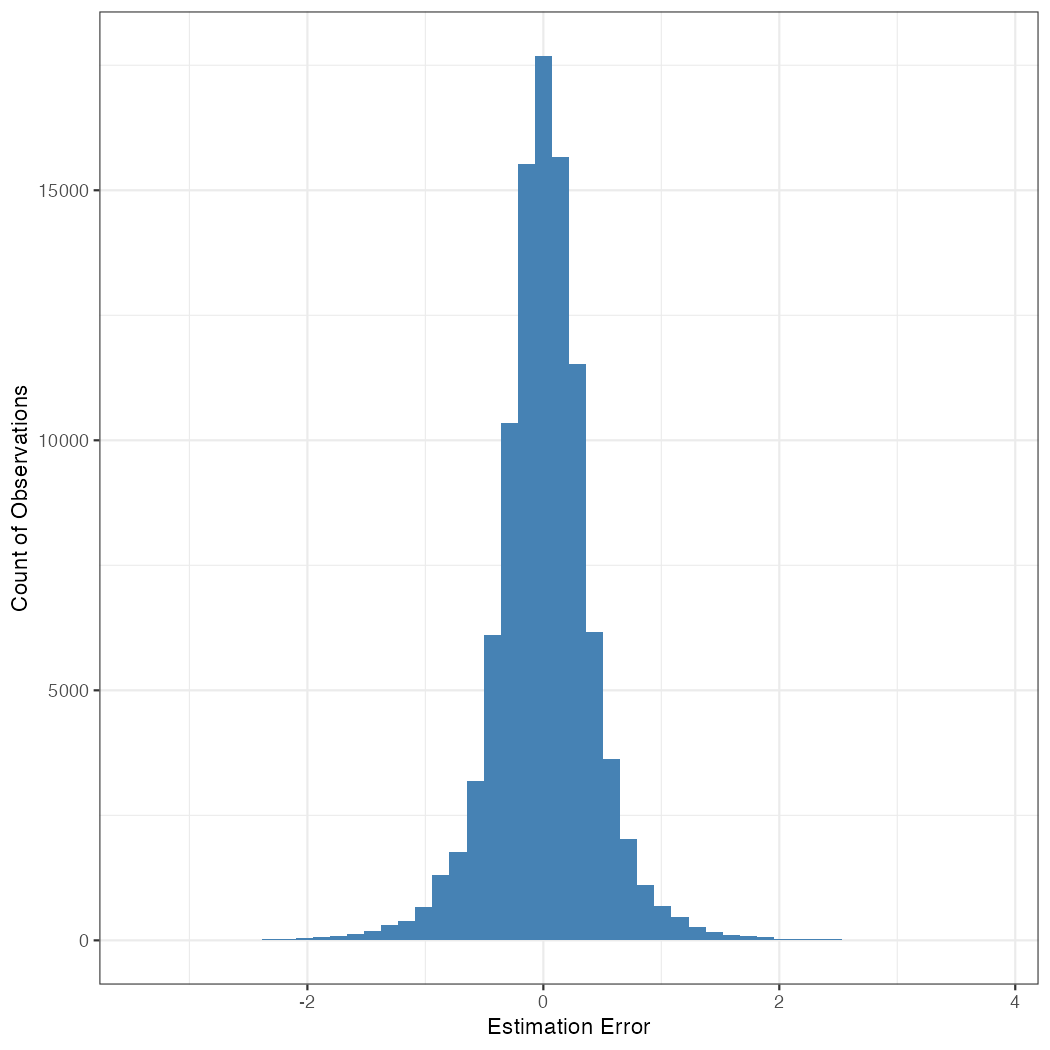}
  \caption{Significant Fixed Effects}
  \label{fig:estimate_against_true_4a}
\end{subfigure}%
\begin{subfigure}{.5\textwidth}
  \centering
  \includegraphics[width=\linewidth]{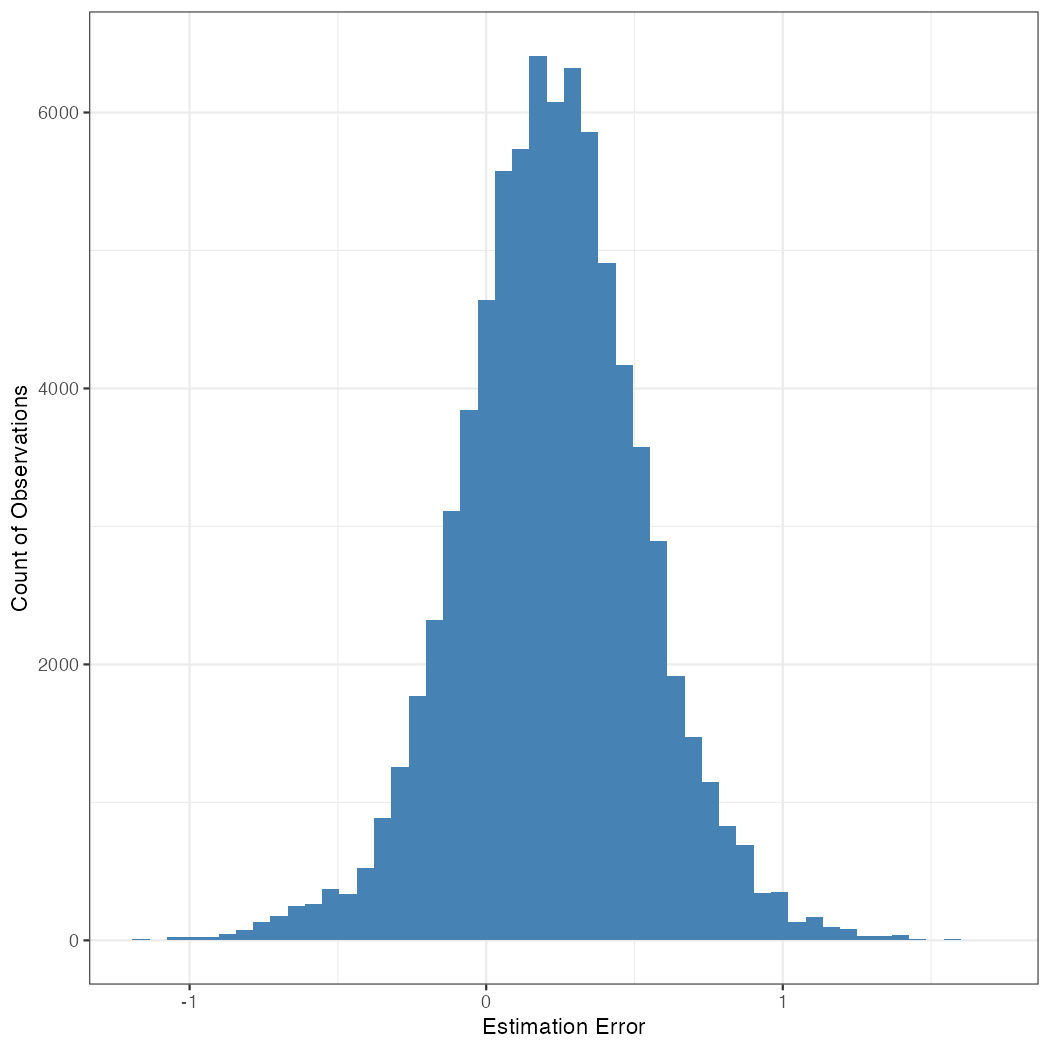}
  \caption{Merged Category Levels}
  \label{fig:estimate_against_true_4b}
\end{subfigure}
\begin{subfigure}{.5\textwidth}
  \centering
  \includegraphics[width=\linewidth]{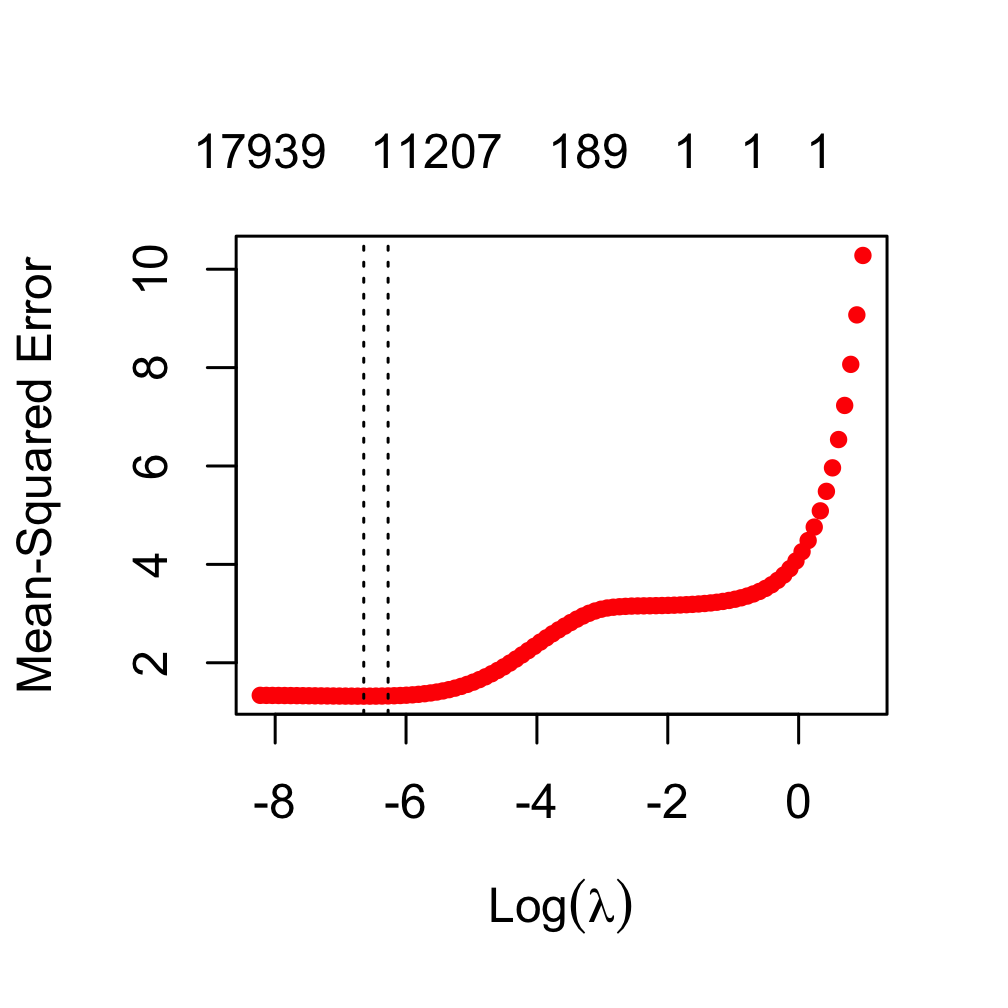}
  \caption{Lambda Selection}
  \label{fig:estimate_against_true_4c}
\end{subfigure}%
\begin{subfigure}{.5\textwidth}
  \centering
  \includegraphics[width=\linewidth]{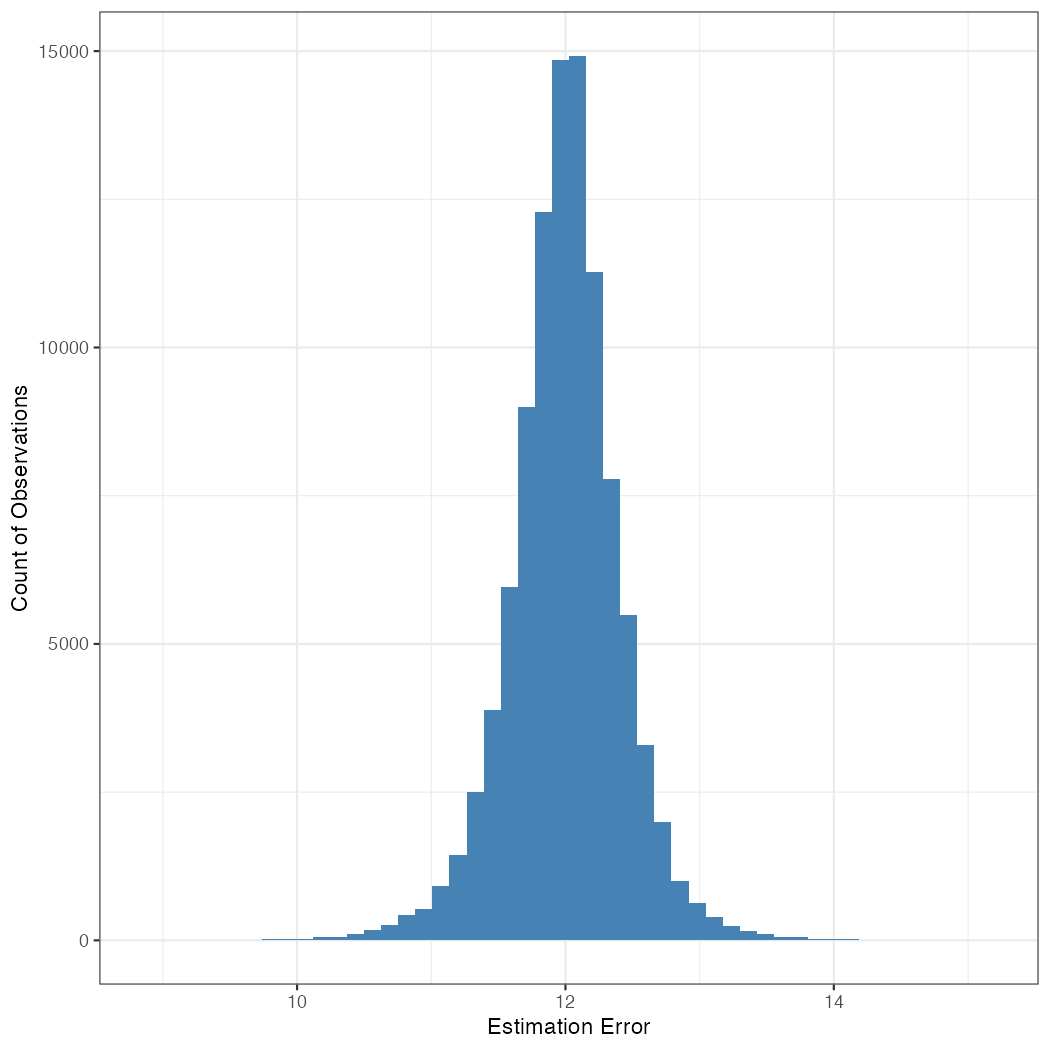}
  \caption{Estimation Errors, LASSO}
  \label{fig:estimate_against_true_4d}
\end{subfigure}
\caption{Simulation 2: Estimation Errors in Regression Analysis}
\label{fig:estimate_against_true_4}
\begin{minipage}{\linewidth}
\medskip
\footnotesize
Note: Estimation accuracy and precision in simulation 2 for the fixed effect coefficients.
\end{minipage}
\end{figure}

Figure \ref{fig:estimate_against_true_3} and Figure
\ref{fig:estimate_against_true_4} characterize estimation errors from
alternative approaches--the former presenting results from simulation 1
and the latter from simulation 2. Subfigures
\ref{fig:estimate_against_true_3a} and
\ref{fig:estimate_against_true_4a} present estimation errors
corresponding to `significant' estimates--which are coefficients that
when tested against the null hypothesis yield a significant result. In
these subfigures, we consider the classic and typical null hypothesis of
the category level having no effect (the true coefficient is 0) as in
practice, the true coefficient is not known and the estimated
coefficients are statistically compared to 0. It is then typical to
report and explain the significant coefficients as in those cases, the
evidence supports a nonzero effect of the category level.

We find that 14,288 fixed effect coefficients (out of 18,468 total
estimated fixed coefficients) are significant at the 5\% level in
simulation 1 and 18,467 fixed effect coefficients are significant in
simulation 2; only one coefficient is nonsignificant in simulation 2.
This illustrates a key concern in this strategy whereby the probability
of a coefficient being significant is the composite of estimation
precision and effect size. Thus, in this strategy, while the reporting
strategy does not bias estimates as each estimate is still derived from
the inclusion of all categorical levels, the reported levels are biased
towards the confluence of lower estimation uncertainty and higher
effective sample size. Thus, for example, Figure
\ref{fig:estimate_against_true_2a} is virtually indistinguishable from
Figure \ref{fig:estimate_against_true_4a} as the histogram differs on
only 1 point estimate--the single coefficient that is nonsignificant at
the 5\% level.

Subfigures \ref{fig:estimate_against_true_3b} and
\ref{fig:estimate_against_true_4b}, we present estimates from an
estimator where infrequent category levels are merged for inclusion with
the intercept, and the model is estimated using the `lm' function in R.
In this case, we observe the mean of the fixed effects is no longer zero
due to a change in the meaning of the fixed effects. The process does
tamp down the variance. Thus, we observe lowered variance at the expense
of bias.

Subfigures \ref{fig:estimate_against_true_3c} and
\ref{fig:estimate_against_true_3d}, as well as
\ref{fig:estimate_against_true_4c} and
\ref{fig:estimate_against_true_4d}, relate to the use of LASSO,
estimated with the `glmnet' function from the `glmnet' package in R,
where the \(\lambda\) hyperparameter is selected through
cross-validation using `cv.glmnet'. The first two subfigures pertain to
simulation 1, and the latter two to simulation 2; the first and third
subfigures describe model performance for various values of \(\lambda\).
The second and fourth subfigures depict the estimation errors when
`lambda.1se', which is the largest value of \(\lambda\) such that the
error is within 1 standard deviation of the cross-validated minimum, is
chosen.

Cross-validated LASSO suggests the inclusion of many categorical
variable levels. In simulation 1, the minimum cross-validated error is
achieved when lambda is 0.001453, and 15,242 levels are included. The
largest value of \(\lambda\) such that the error is within 1 standard
deviation of the cross-validated errors is 0.002109, where 13,784 levels
are included. In simulation 2, the corresponding numbers are 0.001299
and 16,002 for the minimum error, and 0.001884 and 14,920 levels for
error within 1 standard deviation of the minimum. As a reminder, the
data includes 18,468 levels, of which 5,031 were only observed once.
Thus, the use of LASSO does not lead to the exclusion of variables that
are estimated using only 1 observation---in many cases, the estimation
of a fixed effect even from a single observation is critical to
cross-validated model fit, as in its absence, the average of the
excluded fixed effects is a poor substitute for the excluded category
level.

Critically, despite the inclusion of the majority of the categorical
variables in an effort to reduce the variance of estimates, we observe
both bias and considerable imprecision in subfigures
\ref{fig:estimate_against_true_3d} and
\ref{fig:estimate_against_true_4d} that depict the estimation errors
when `lambda.1se', which is the largest value of \(\lambda\) such that
the error is within 1 standard deviation of the cross-validated minimum,
is chosen. This illustrates the dilemma such data presents to the
researcher: including all categorical levels leads to imprecision, but
merging or dropping levels through the use of an ad-hoc strategy or
through principled discovery using cross-validated LASSO is also
ineffective at mitigating instability.

\begin{figure}[htbp]
\centering
\begin{subfigure}{.5\textwidth}
  \centering
  \includegraphics[width=\linewidth]{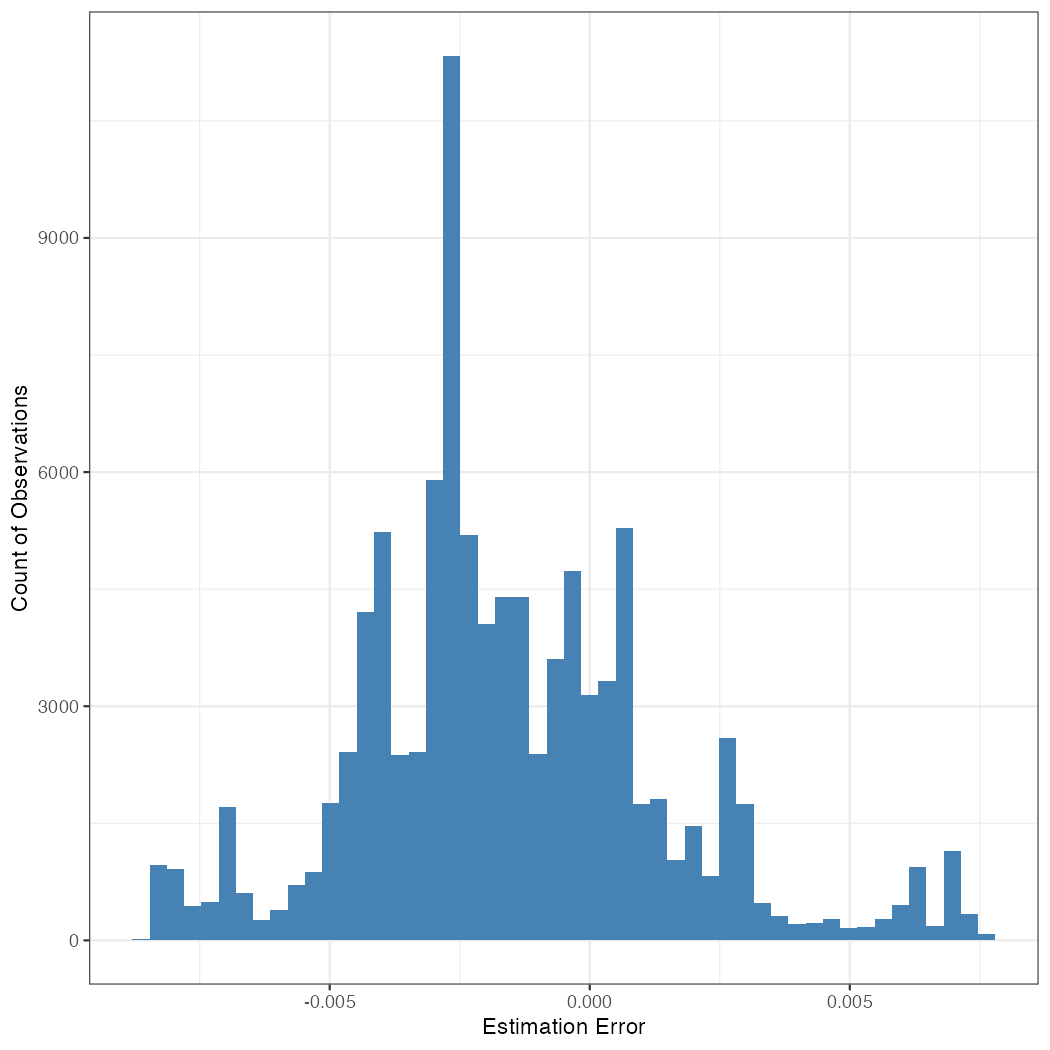}
  \caption{Simulation 1: Histogram of Estimation Errors}
  \label{fig:estimate_our_modela}
\end{subfigure}%
\begin{subfigure}{.5\textwidth}
  \centering
  \includegraphics[width=\linewidth]{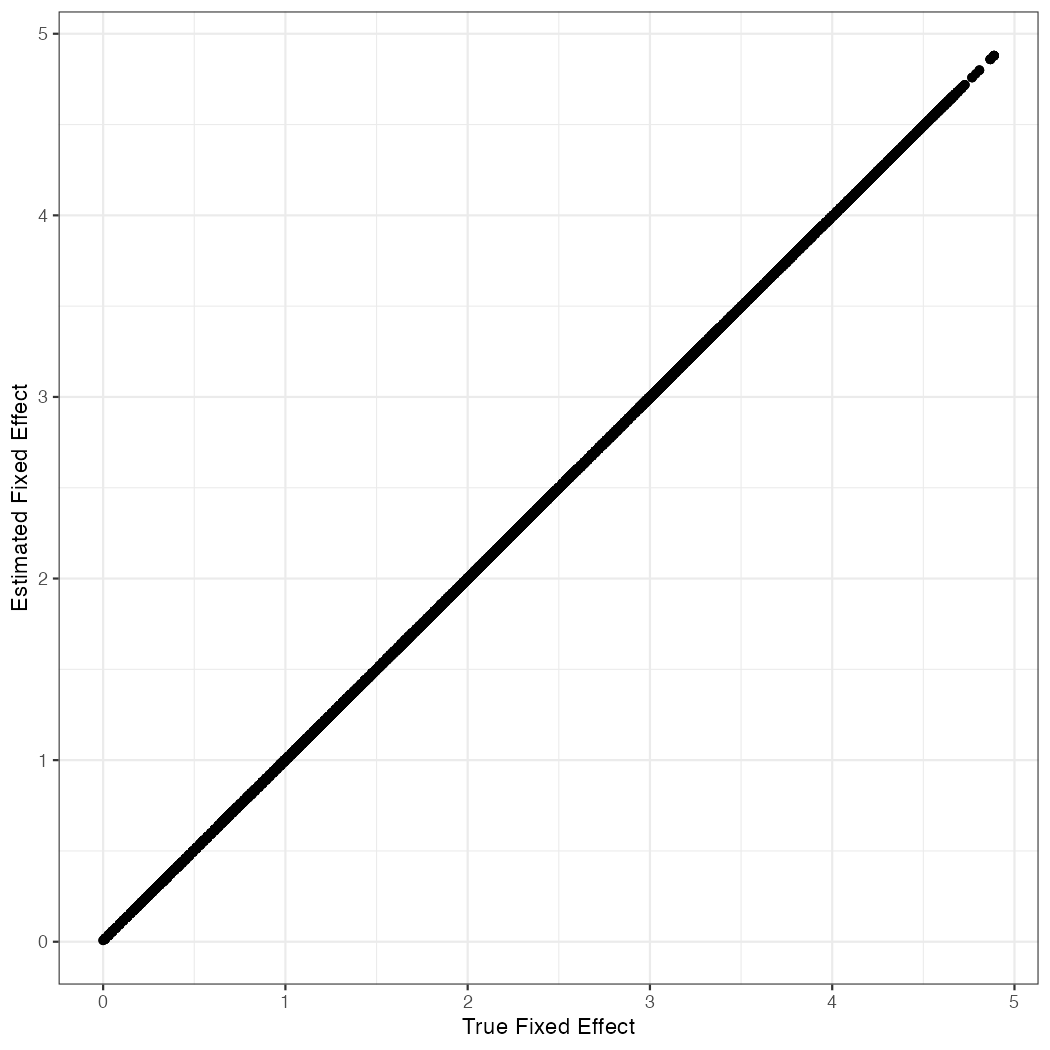}
  \caption{Simulation 1: Inferred Fixed Effect vs. True Fixed Effect}
  \label{fig:estimate_our_modelb}
\end{subfigure}
\begin{subfigure}{.5\textwidth}
  \centering
  \includegraphics[width=\linewidth]{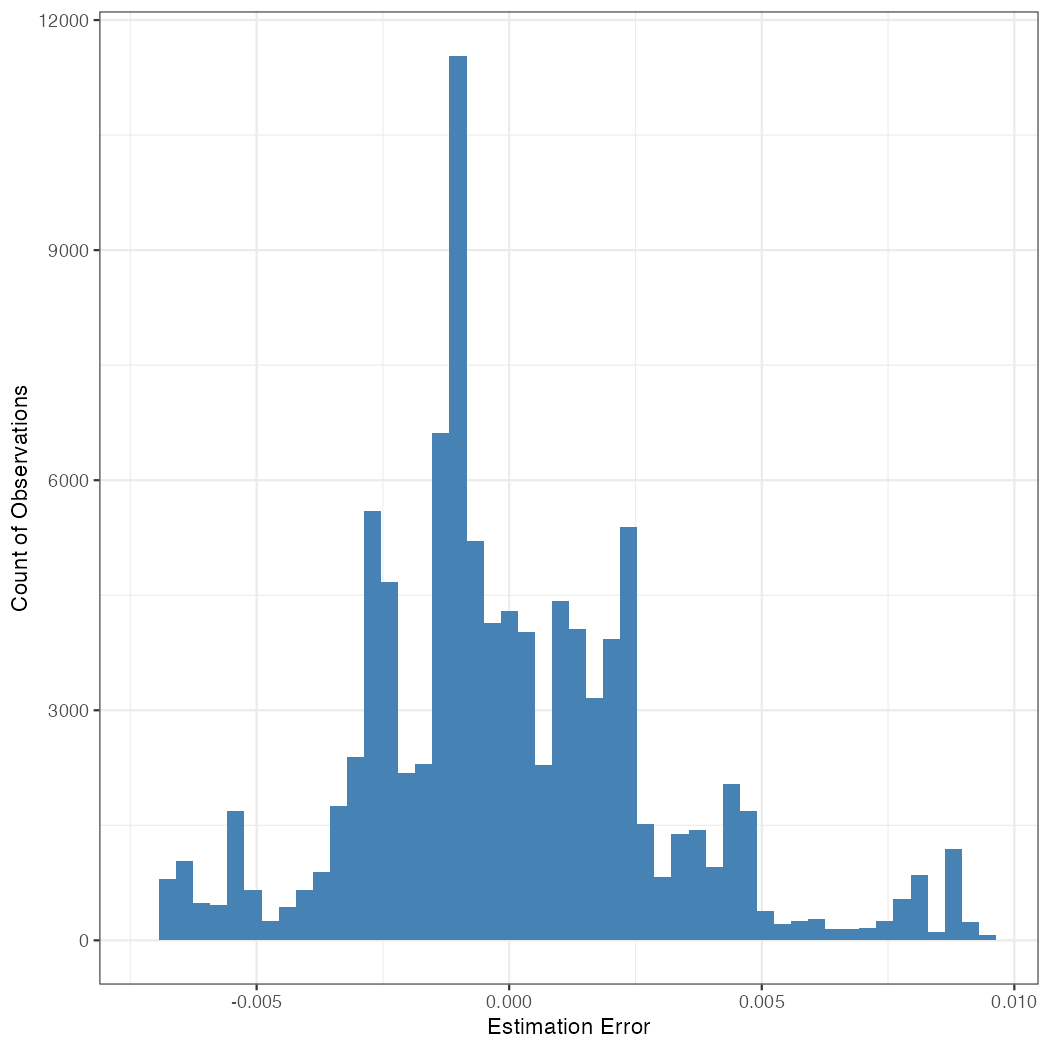}
  \caption{Simulation 2: Histogram of Estimation Errors}
  \label{fig:estimate_our_modelc}
\end{subfigure}%
\begin{subfigure}{.5\textwidth}
  \centering
  \includegraphics[width=\linewidth]{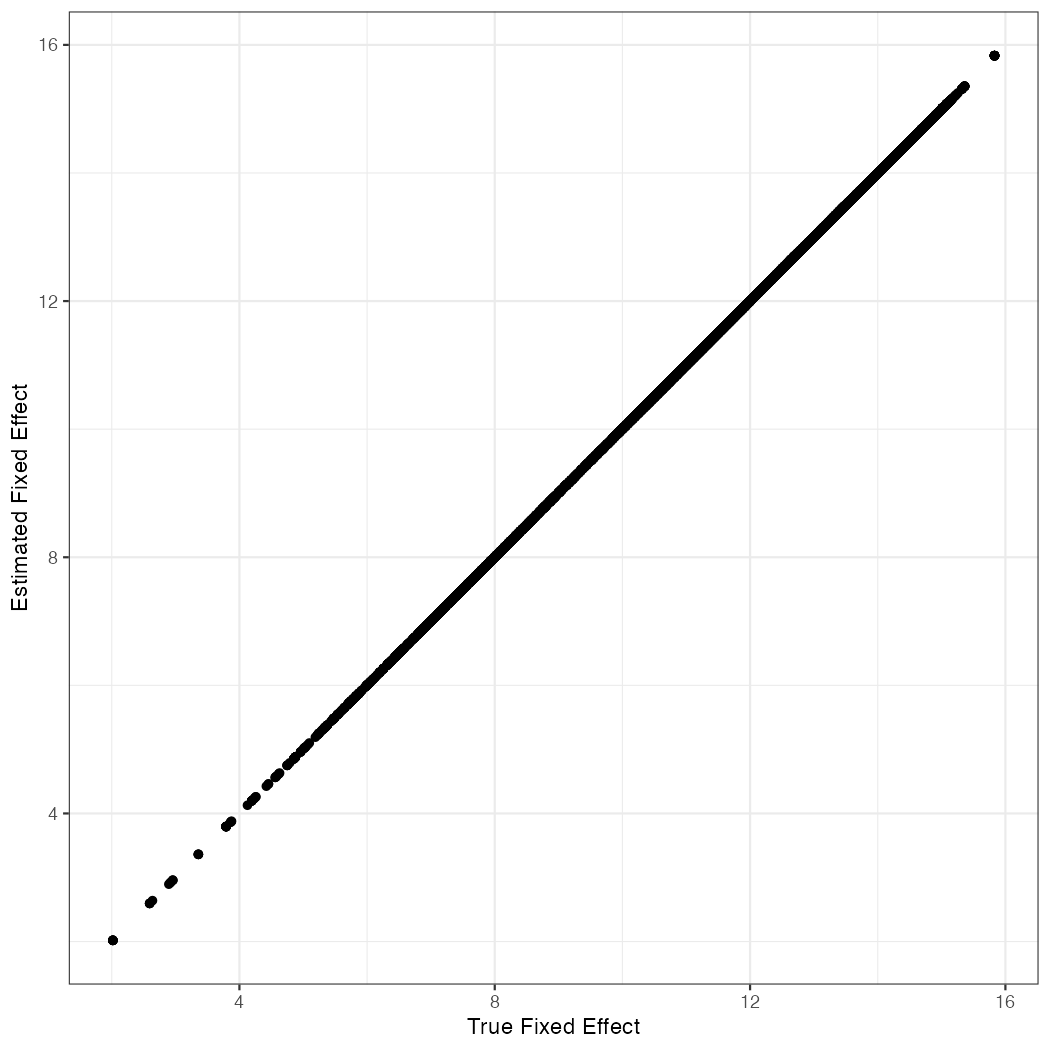}
  \caption{Simulation 2: Inferred Fixed Effect vs. True Fixed Effect}
  \label{fig:estimate_our_modeld}
\end{subfigure}
\caption{Proposed Estimator Performance}
\label{fig:estimate_our_model}
\begin{minipage}{\linewidth}
\medskip
\footnotesize
Note: Estimation accuracy and precision using our proposed estimator for the fixed effect coefficients.
\end{minipage}
\end{figure}

Figure \ref{fig:estimate_our_model} presents estimates from our proposed
approach in both simulations. The figure comprises four subfigures of
which subfigures \ref{fig:estimate_our_modela} and
\ref{fig:estimate_our_modelb} address simulation 1, and subfigures
\ref{fig:estimate_our_modelc} and \ref{fig:estimate_our_modeld} address
simulation 2. The first and third subfigures are histograms of the
estimation error and the second and fourth subfigures plot the estimate
against the true coefficient. In all cases, our proposed approach yields
analysis that is more accurate and more precise as it relates the fixed
effect to its determinants. The imposed structure provides the model
with both the statistical scaffolding needed to ensure precise
estimation, and a means to capture the variation in the fixed effects.

\hypertarget{application-direct-to-consumer-sales}{%
\section{Application: Direct-to-Consumer
Sales}\label{application-direct-to-consumer-sales}}

We analyze data from an online apparel provider that was a pioneer in
direct-to-consumer sales. Our data covers a period from the early days
of internet retailing when the firm was the market leader, facing little
to no competition from similar firms. Therefore, our data uniquely
tracks the evolution of both the firm and the industry during this
pivotal period. Although the identity of the data-providing firm cannot
be disclosed, it is noteworthy that the firm launched waves of new
apparel products sold exclusively to consumers through their platform.
For the purposes of our analysis, we operate under the assumption that
the firm primarily sold summer apparel, which serves as an illustrative
example of their product category. Given that the firm was founded on
the East Coast and initially found success in both the East Coast and
Midwest regions, we anticipate, contrary to our simulations, observing
the influence of longitude on fixed effects.

Our overarching goal is to conduct demand analysis, aiming to measure
demand at the zip code level. Specificity in such analyses is crucial as
it informs managerial decisions, such as the locations for physical
stores. In this particular case, the firm initially lacked a
brick-and-mortar presence but later expanded offline. While specific
details about store openings are confidential and not directly traceable
in this study to protect the company's identity, we can confirm that the
selected cities for physical stores align with the `hot spots'
identified in our data---areas showing the highest demand for the
company's products.

\hypertarget{data}{%
\subsection{Data}\label{data}}

\begin{figure}[htbp]
\centering
\includegraphics[width=0.75\linewidth]{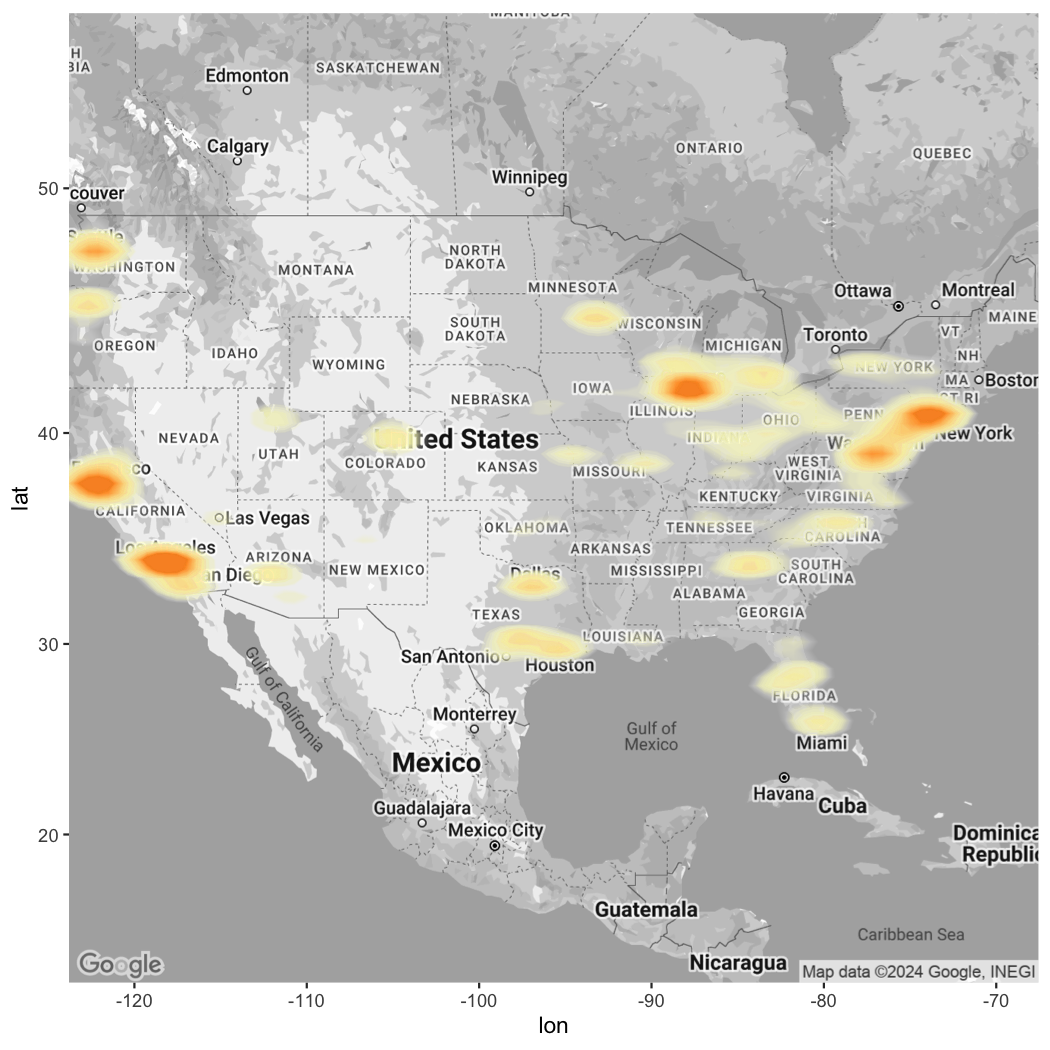}
\caption{Heatmap of Order Volume in the Contiguous United States}
\label{fig:map_field}
\begin{minipage}{\linewidth}
\medskip
\footnotesize
Note: Color and opacity represent order volume.
\end{minipage}
\end{figure}

The data pertains to orders from the contiguous United States. Figure
\ref{fig:map_field} plots a heatmap of order frequency by zip code on a
map of the United States, where a darker color corresponds to higher
order volume for a zip code. We see that the order volume is anchored in
a few major cities such as Seattle, San Francisco, Los Angeles, Chicago,
New York, and Washington DC, but otherwise dispersed throughout the
United States.

\begin{figure}[htbp]
\centering
\includegraphics[width=0.5\linewidth]{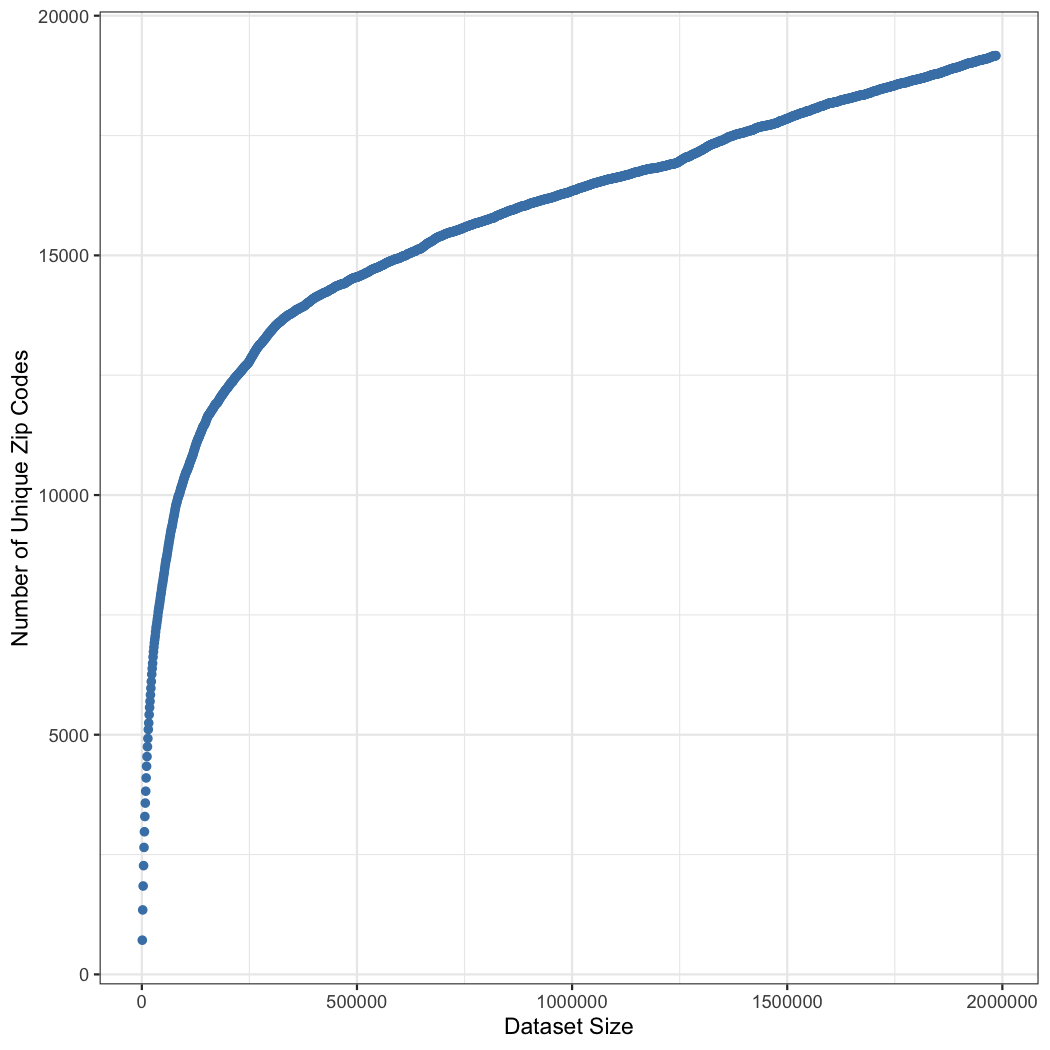}
\caption{Number of Unique Categories vs. Sample Size}
\label{fig:num_categories_field}
\begin{minipage}{\linewidth}
\medskip
\footnotesize
Note: On the y-axis is the number of unique zip codes. On the x-axis is the sample size.
\end{minipage}
\end{figure}

We present the following figures to illustrate the key modeling
challenge. Figure \ref{fig:num_categories_field} plots the number of
unique zip codes for every 1,000 observations in the data, starting from
the inception of the firm and ordered according to when the observations
occurred. This figure demonstrates the number of unique zip codes that
the firm might expect to encounter in a data sample of a specified size
if the data were collected from the beginning of the firm's operation.
Specifically, our complete dataset comprises 1,985,365 observations.
Therefore, the figure presents the number of unique zip codes for nearly
2,000 data samples, illustrating the growth in the number of zip codes,
which is central to the data challenge that we seek to address in this
paper.

We observe a steep increase in the number of unique zip codes in the
data during the initial quarter of the data, covering the first 500,000
observations. While the rate at which additional zip codes are observed
in the data diminishes with the accrual of data, throughout our entire
dataset of almost 2 million observations, we find that the number of zip
codes does not level off and continues to increase with the data. Thus,
in a classical model where zip codes are introduced in the specification
as a categorical variable, the number of levels would remain a function
of sample data points for the duration of our dataset.

\begin{figure}[htbp]
\centering
\begin{subfigure}{.5\textwidth}
  \centering
  \includegraphics[width=\linewidth]{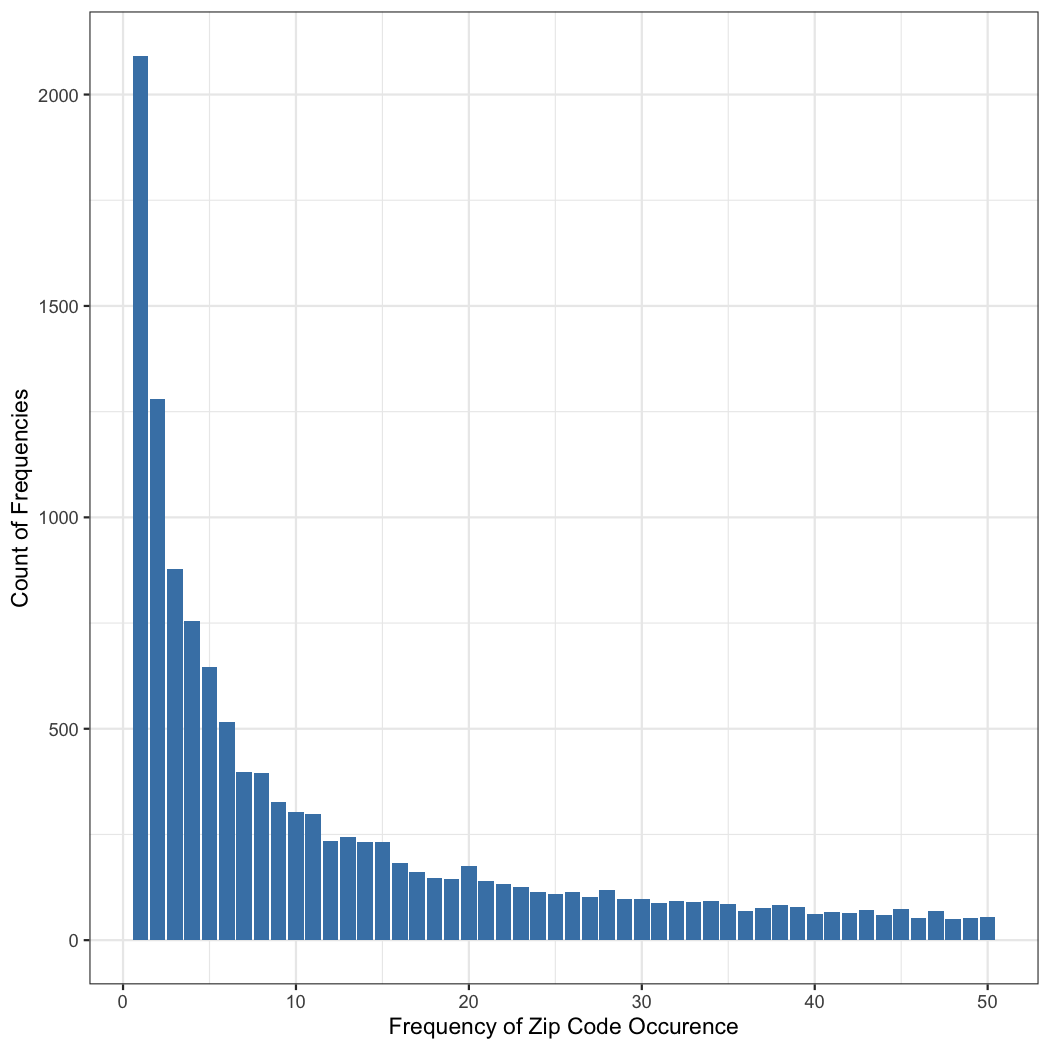}
  \caption{Distribution of Least Frequent Zip Codes}
  \label{fig:frequency_obs_a}
\end{subfigure}%
\begin{subfigure}{.5\textwidth}
  \centering
  \includegraphics[width=\linewidth]{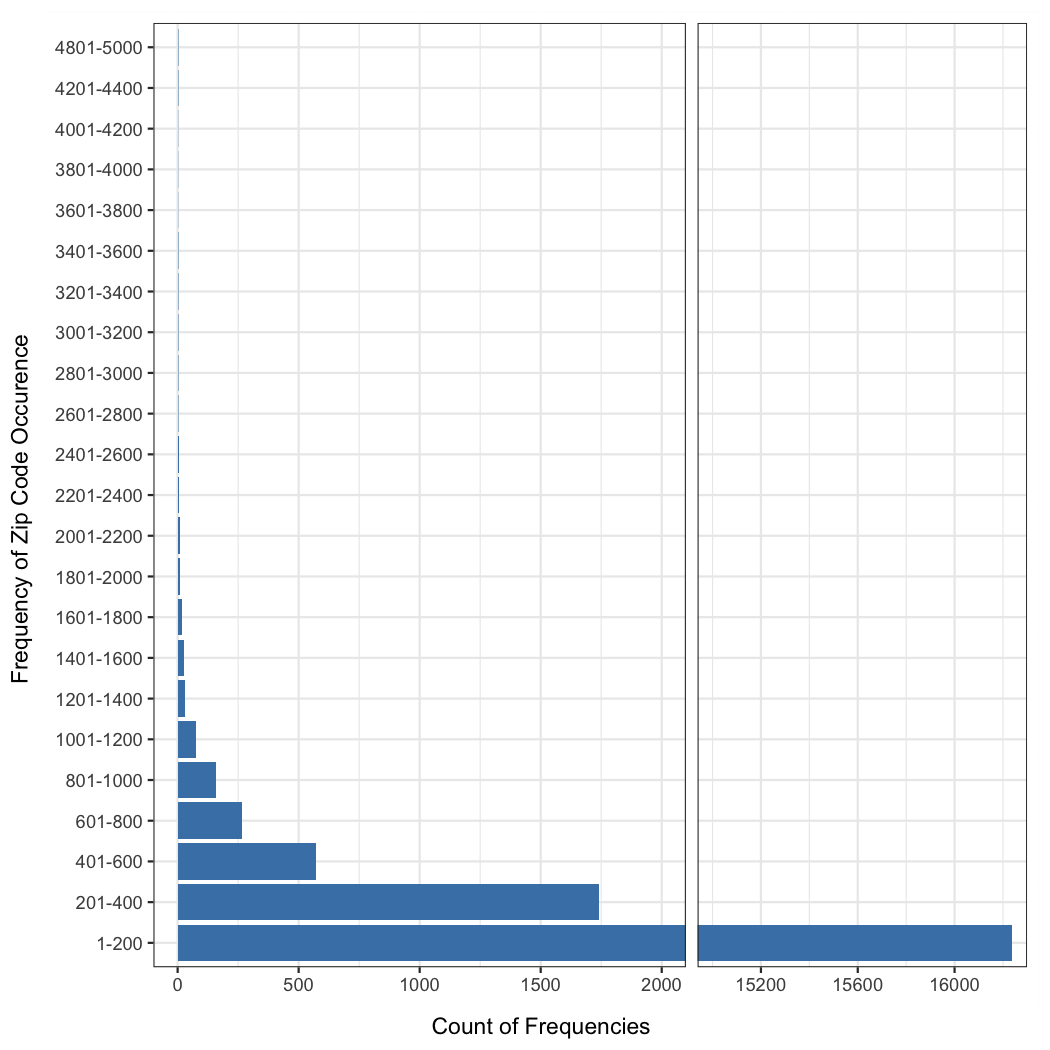}
  \caption{Distribution of All Frequent Zip Codes}
  \label{fig:frequency_obs_b}
\end{subfigure}
\caption{Distribution of Zip Code Frequencies in Complete Sample}
\label{fig:frequency_obs}
\begin{minipage}{\linewidth}
\medskip
\footnotesize
Note: Frequency with which a zip code occurred in the field data and the count of zip codes that occurred with a given frequency.
\end{minipage}
\end{figure}

Analogous to Figure \ref{fig:frequency} in our simulation data, Figure
\ref{fig:frequency_obs} illustrates the frequency table of the zip codes
in our field data. The figure is split into two subfigures for improved
presentation: the top panel (subfigure \ref{fig:frequency_obs_a})
presents the count of frequencies on the y-axis and the frequency on the
x-axis for the least common zip codes; the bottom panel (subfigure
\ref{fig:frequency_obs_b}) presents similar statistics with the axes
reversed for all frequencies of all zip codes.

The zip codes (the focal independent variable in our study) exhibit a
power-law distribution, whereby a few zip codes have many observations
(the most frequent zip code in our data---Chicago, IL, 60657---has 5,758
observations), and 3,372 zip codes have either one or two corresponding
observations. In total, the data pertains to 19,170 zip codes, which
represents only about two-thirds of the zip codes in the United States.
Therefore, the sparsity concerns we highlight are likely to persist even
if a much larger data sample were collected.

Including 20,000 zip codes in an econometric model, and estimating on 2
million data points, would likely be impractical for most researchers.
For instance, with 20,000 indicator variables and 2 million
observations, the design matrix would require approximately 320 GB of
memory just for storage---estimation would necessitate even more memory.
These data demands would scale supralinearly with the inclusion of more
data as in addition to adding rows to the design matrix, the addition of
data is likely to lead to the addition of indicator variables through
the addition of novel category levels (i.e., zip codes).

\begin{figure}[htbp]
\centering
\includegraphics[width=0.5\linewidth]{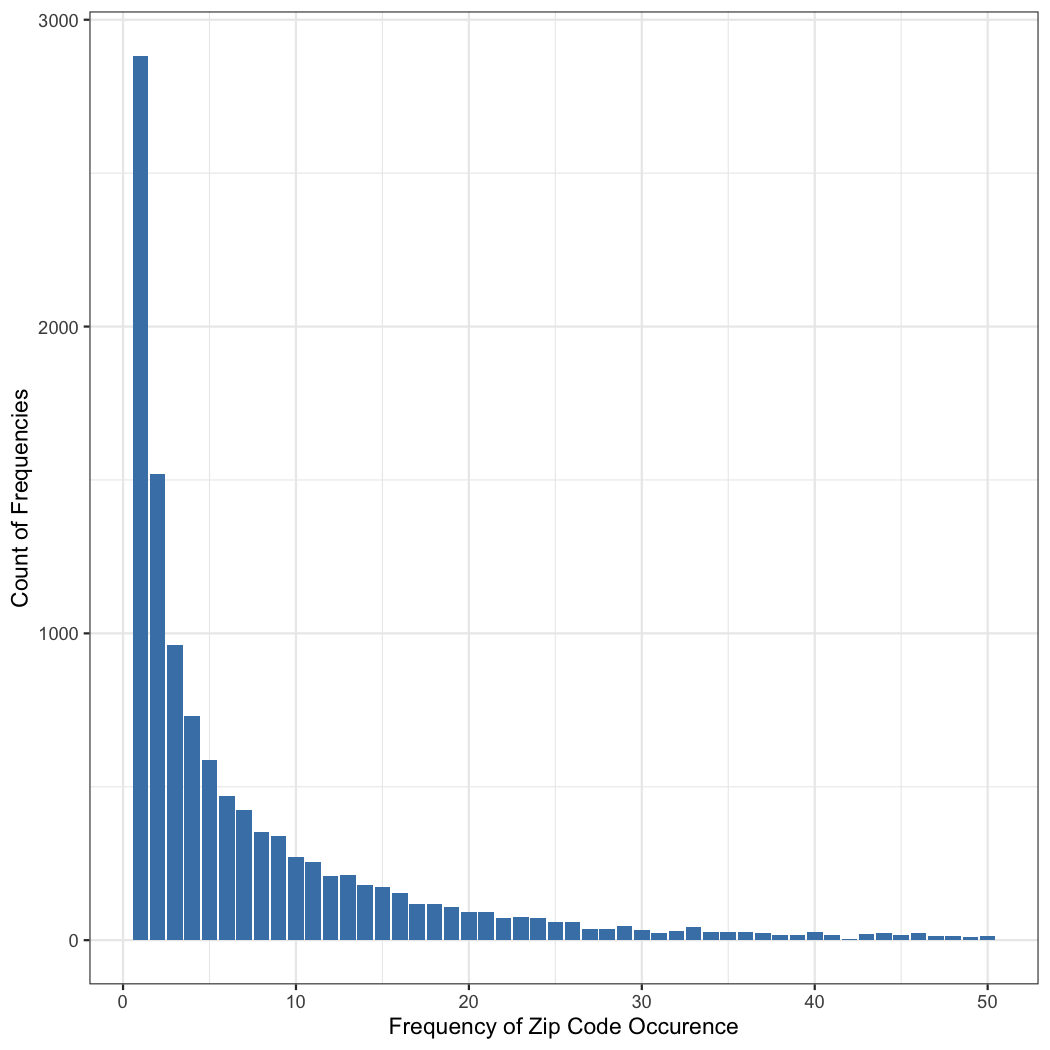}
\caption{Distribution of Zip Code Frequencies in Data Subsample}
\label{fig:frequency_obs_sub}
\begin{minipage}{\linewidth}
\medskip
\footnotesize
Note: The x-axis represents the frequency with which a zip code occurred in our field data. The y-axis shows the count of zip codes that occurred with each given frequency.
\end{minipage}
\end{figure}

Therefore, we constructed a random sample of 100,000 observations from
the complete dataset for estimation purposes. Figure
\ref{fig:frequency_obs_sub} presents the frequency table of zip codes in
this data subsample. As expected, a power-law distribution of zip codes
is observed, with 5,272 zip codes appearing two or fewer times in the
subsample and one zip code appearing 269 times. It's noteworthy that the
sparsity patterns in our field data subsample are less pronounced than
those observed in our simulations. Specifically, in our simulations,
18,468 zip codes were selected at least once, whereas in our field data
subsample, 11,411 zip codes were selected at least once. This
discrepancy suggests that the dispersion of orders across zip codes in
the field data is not entirely random. Likely, the firm's products are
more popular in certain zip codes, leading to repeat orders.
Nonetheless, the field data still illustrates the power law distribution
that is central to the modeling challenge addressed in our paper.

\hypertarget{model-specification}{%
\subsection{Model Specification}\label{model-specification}}

Our dependent variable is revenue per order (inclusive of any shipping
and handling charges). As a focal covariate, we consider the use of a
coupon, a common practice in internet retailing that significantly
lowers the order value and drives order growth. We incorporate fixed
effects for zip codes, to delineate the firm's geographic expansion and
include dummy variables for both month and year to account for
seasonality and to capture the growth in the firm's product portfolio
and its appeal over time. Thus, we specify the model as follows:

\[sales_{it} = \alpha + \beta_p \text{coupon}_{it} + \sum_z \beta_z D_{iz} + \sum_m \beta_m D_{mt} + \sum_y \beta_y D_{yt} + \epsilon_{it},\]

where:

\begin{itemize}
\tightlist
\item
  \(sales_{it}\) is revenue inclusive of shipping and handling.
  \(\alpha\), \(\beta_p\), \(\{\beta_z\}_z\), \(\{\beta_m\}_m\), and
  \(\{\beta_y\}_y\) denote the intercept, the coefficient on the coupon,
  and the fixed effects for zip code, month of purchase, and year of
  purchase, respectively. \(t\) represents the time index, and \(i\)
  signifies the consumer index.
\item
  \(\text{coupon}_{it}\) is an indicator variable that equals one if
  consumer \(i\) utilized a coupon at time \(t\).
\item
  \(D_{iz}\) is an indicator variable that equals one if consumer \(i\)
  resides in zip code \(z\).
\item
  \(D_{mt}\) and \(D_{yt}\) are indicator variables for the month and
  year of purchase, respectively, equaling one if \(t\) corresponds to
  month \(m\) and year \(y\).
\item
  \(\epsilon_{it}\) is the error term, assumed to have a mean of zero.
\end{itemize}

To predict the fixed effects associated with each zip code, we consider
the latitude, longitude, and elevation of each zip code. Given the
anonymization of consumer addresses to the zip code level in our
dataset, we utilize the centroid of each zip code to derive these
auxiliary variables.

We introduce two forms of additional information to enhance our model's
predictive accuracy for zip code fixed effects. First, we incorporate
the encoding from OpenAI's `text-embedding-3-large' model of each zip
code's name and address (e.g., `10001 New York, NY'). This approach
recognizes that distinct zip codes, even those at similar elevations,
can exhibit cultural differences that may influence demand. For
instance, demand for summer apparel in Berkeley, CA, might differ from
that in San Francisco, CA, despite their geographical proximity. Given
the extensive data features provided by both structured data points
(latitude, longitude, and altitude) and the LLM capture of the town/city
name, direct inclusion of the LLM encoding in the statistical model is
impractical. Therefore, we employ PCA for rank reduction, allowing us to
retain essential descriptors while excluding irrelevant data features
related to inter-zip code differences.

Second, we utilize a series of variables describing the geographical and
socio-economic characteristics of each zip code as additional predictors
of the fixed effect. Here, we replace the unstructured LLM encoding
based on the zip code and associated address with structured information
such as median home value, which also reflects systematic differences
among zip codes. To assess the extent of overlap in their informative
content and examine the extent to which these variables provide
complementary insights into the fixed effects associated with each zip
code, we specify a model incorporating both structured and unstructured
information.

Table \ref{tab:summary_statistics} presents key descriptive statistics
of the data for the mainland states of the United States; we exclude
observations related to international sales and those to Hawaii and
Alaska to maintain a contiguous definition of zip codes.

\begin{table}[!htbp] \centering 
  \caption{Descriptive statistics} 
  \label{tab:summary_statistics} 
\begin{tabular}{@{\extracolsep{5pt}}lccccc} 
\\[-1.8ex]\hline 
\hline \\[-1.8ex] 
Variable & \multicolumn{1}{c}{N} & \multicolumn{1}{c}{Mean} & \multicolumn{1}{c}{St. Dev.} & \multicolumn{1}{c}{Min} & \multicolumn{1}{c}{Max} \\ 
\hline \\[-1.8ex] 
Revenue Per Order & 1,985,365 & 35.966 & 40.899 & 0.010 & 14,402.380 \\ 
Coupon & 1,985,365 & 0.084 & 0.278 & 0 & 1 \\ 
\hline \\[-1.8ex] 
Latitude & 1,985,365 & 37.895 & 4.971 & 24.600 & 48.990 \\ 
Longitude & 1,985,365 & $-$96.830 & 17.413 & $-$124.500 & $-$71.940 \\ 
Elevation & 1,985,365 & 250.328 & 374.123 & $-$858 & 3,870 \\ 
\hline \\[-1.8ex] 
Housing Units & 1,985,365 & 14,263.680 & 7,439.505 & 16 & 47,617 \\ 
Land Area & 1,985,365 & 35.693 & 91.585 & 0.020 & 3,911.400 \\ 
Median Home Value & 1,985,365 & 328,507.200 & 223,369.500 & 10,200 & 1,000,001 \\ 
Median Household Income & 1,985,365 & 66,062.560 & 26,331.730 & 2,499 & 250,001 \\ 
Occupied Housing Units & 1,985,365 & 13,118.470 & 6,857.844 & 12 & 44,432 \\ 
Population Density & 1,985,365 & 6,703.375 & 13,711.620 & 0 & 143,683 \\ 
Water Area & 1,985,365 & 1.040 & 4.171 & 0.000 & 255.390 \\ 
\hline \\[-1.8ex] 
\end{tabular} 
\end{table}

\hypertarget{results-1}{%
\subsection{Results}\label{results-1}}

Initially, we present findings from regressions conducted on our random
subsample of 100,000 observations across four distinct scenarios, where
the zip code variable is specified at different levels of aggregation.
This approach leverages the hierarchical nature of zip codes, with the
first digit representing a broad geographic grouping of states, two
digits representing a sectional center facility (SCF, a major mail
processing center within a region), three digits designating a more
specific area within the SCF, and the full zip code pinpointing a
specific delivery area, such as a neighborhood or a group of streets.
While this hierarchy may not always effectively capture consumption
pattern differences---for example, Buffalo, NY, may have more distinct
consumption patterns from Manhattan, NY, than Princeton, NJ, a city much
closer to Manhattan---it offers a structured approach to the inclusion
of categorical variables. This approach is particularly useful in
practice when faced with the challenges of managing many unique levels
and data that might otherwise overwhelm the model.

Table \ref{tab:regressions1} presents our findings. Columns 1, 2, 3, and
4 present results from the inclusion of 1-digit, 2-digit, 3-digit, and
5-digit zip code levels, respectively. The first two models are
relatively stable, with 8 and 89 fixed effects estimated from 100,000
observations. The third model is also somewhat stable, as more than 100
observations correspond to each zip code on average. However, in 14
cases, only 1 observation supports inference, and in 13 cases, only 2
observations support inference. Thus, in the third model, the effects of
category sparsity begin to emerge, even though the data is sufficiently
large to support the estimation of many of the included zip code fixed
effects.

The fourth model is saturated, as the data includes 11,411 unique zip
codes. For instance, in 2,961 cases, inference is supported by 1
observation; in 1,450 cases, inference is supported by 2 observations;
and in 986 cases, inference is supported by 3 observations.
Consequently, the models' adjusted R-squared, which is 0.028 and 0.029
for the first two models that are not saturated, increases to 0.032 for
the third model that is slightly saturated and 0.151 for the fourth
model, which is highly saturated. This saturation implies that the
residuals corresponding to observations where the categories are sparse
(i.e., only one or a few observations correspond to the category) are
fit with little to no error. In these cases, the fixed effect suffices
to perfectly or almost perfectly explain the observations, improving the
fit statistic drastically. Our knowledge that the first 2 models are
relatively less saturated suggests that an R-squared of about 0.029 is a
more accurate reflection of the extent to which zip codes explain the
variance in order values across geography.

\begin{table}[!htbp] \centering 
  \caption{Zip Code-level Inference, Fixed Effects} 
  \label{tab:regressions1}
\begin{tabular}{@{\extracolsep{5pt}}lcccc} 
\\[-1.8ex]\hline 
\hline \\[-1.8ex] 
 & \multicolumn{4}{c}{Dependent variable: Revenue Per Order} \\ 
\cline{2-5} 
\\[-1.8ex] & (1) & (2) & (3) & (4)\\ 
\hline \\[-1.8ex] 
 Coupon & $-$20.239$^{***}$ & $-$20.215$^{***}$ & $-$20.292$^{***}$ & $-$20.537$^{***}$ \\ 
 Constant & 37.971$^{***}$ & 41.020$^{***}$ & 41.143$^{***}$ & 46.224$^{***}$ \\ 
\hline \\[-1.8ex] 
\# 1-Digit Zip Code & 9 & 0 & 0 & 0 \\ 
\# 2-Digit Zip Code & 0 & 90 & 0 & 0 \\ 
\# 3-Digit Zip Code & 0 & 0 & 787 & 0 \\ 
\# 5-Digit Zip Code & 0 & 0 & 0 & 11411 \\ 
\hline \\[-1.8ex] 
Observations & 100,000 & 100,000 & 100,000 & 100,000 \\ 
R$^{2}$ & 0.028 & 0.030 & 0.039 & 0.151 \\ 
Adjusted R$^{2}$ & 0.028 & 0.029 & 0.032 & 0.043 \\
Residual Std. Error & 33.980 & 33.956 & 33.907 & 33.715 \\
 & (df = 99990) & (df = 99909) & 
(df = 99217) & (df = 88726) \\
F Statistic & 316.956$^{***}$ & 34.233$^{***}$ & 5.210$^{***}$ & 1.396$^{***}$ \\ 
 & (df = 9; 99990) & (df = 90; 99909) & (df = 782; 99217) & (df = 11273; 88726) \\ 
\hline 
\hline \\[-1.8ex] 
\textit{Note:}  & \multicolumn{4}{r}{$^{*}$p$<$0.1; $^{**}$p$<$0.05; $^{***}$p$<$0.01} \\ 
\end{tabular} 
\end{table}

We used a subsample in this analysis due to computational stability and
cost concerns. Running a regression with 100,000 observations and
five-digit zip code fixed effects using `lm' in R took approximately 4
hours on a multicore state-of-the-art workstation (Apple Studio M2
Ultra). Given that computational costs in regression analysis tend to
scale cubically, we estimate that even if the workstation had access to
sufficient working memory, the computational time would be
cost-prohibitive, potentially requiring upwards of 32,000 hours for our
entire dataset.

Moreover, while routines have been developed for fast fixed effect
estimation, such as iteratively reweighted least squares
(\protect\hyperlink{ref-hinz2019separating}{Hinz et al. 2019}), these
methods rely on distributional assumptions about the link function. In
contrast, identification in the regression model only requires
specifying the orthogonality of the residuals with respect to the
regressors and a convergence of the second moments to a positive
semi-definite matrix. Although a clever pairing of optimization
methodology and specific econometric models may yield a computationally
stable estimation routine in certain cases, estimation in the general
case remains computationally challenging. This computational hurdle, in
addition to the statistical concerns raised by the saturation of degrees
of freedom due to the inclusion of many fixed effect parameters (as seen
in Models 3 and 4), further justifies the development of our model.

Table \ref{tab:regressions2} summarizes estimates when fixed effects are
projected to a lower dimensional space. We specify four models. The
first model includes only the latitude, longitude, and elevation of the
zip code. This model corresponds to our simulations and represents a
baseline explanation of the variation in demand across zip codes. Model
2 includes a set of key economic descriptors that track cultural
(e.g.~population density tracks rural vs.~urban), economic (e.g.,
household), geographic (e.g.~land area), and social (e.g., housing
density) factors that may explain differences in demand.

\begin{table}[!htbp] \centering 
  \caption{Zip Code-level Inference, Proposed Approach} 
  \label{tab:regressions2}
\begin{tabular}{@{\extracolsep{5pt}}lcccc} 
\\[-1.8ex]\hline 
\hline \\[-1.8ex] 
 & \multicolumn{4}{c}{Dependent variable: Revenue Per Order} \\ 
\cline{2-5} 
\\[-1.8ex] & (1) & (2) & (3) & (4)\\ 
\hline \\[-1.8ex] 
 Coupon & $-$21.431$^{***}$ & $-$21.408$^{***}$ & $-$21.432$^{***}$ & $-$21.404$^{***}$ \\ 
\hline \\[-1.8ex] 
 Latitude & $-$0.045$^{***}$ & $-$0.054$^{***}$ &  & $-$0.155$^{***}$ \\ 
 Longitude & $-$0.046$^{***}$ & $-$0.027$^{***}$ &  & $-$0.085$^{***}$ \\ 
 Elevation & $-$0.001$^{***}$ & $-$0.0001$^{*}$ &  & $-$0.0001 \\ 
\hline \\[-1.8ex] 
 Home Value &  & 0.00001$^{***}$ &  & 0.00000$^{***}$ \\ 
 Household Income &  & $-$0.00002$^{***}$ &  & $-$0.00002$^{***}$ \\ 
 Housing Units (HU) &  & 0.0005$^{***}$ &  & 0.0004$^{***}$ \\ 
 Land Area &  & 0.001$^{***}$ &  & 0.002$^{***}$ \\ 
 Occupied HU &  & $-$0.001$^{***}$ &  & $-$0.0005$^{***}$ \\ 
 Population Density &  & 0.00005$^{***}$ &  & 0.00002$^{***}$ \\ 
 Water &  & 0.081$^{***}$ &  & 0.080$^{***}$ \\
\hline \\[-1.8ex] 
 Constant & 35.246$^{***}$ & 36.460$^{***}$ & 36.103$^{***}$ & 30.817$^{***}$ \\ 
\hline \\[-1.8ex] 
OpenAI Encoding & No & No & Yes & Yes \\ 
\hline \\[-1.8ex] 
Observations & 1,985,365 & 1,985,365 & 1,985,365 & 1,985,365 \\ 
R$^{2}$ & 0.022 & 0.023 & 0.023 & 0.024 \\ 
Adjusted R$^{2}$ & 0.022 & 0.023 & 0.023 & 0.024 \\ 
Residual Std. Error & 40.456 & 40.428 & 40.418 & 40.410 \\
 & (df = 1985360) & (df = 1985353) & (df = 1985331) & (df = 1985321) \\
F Statistic & 10,947.300$^{***}$ & 4,231.140$^{***}$ & 1,441.538$^{***}$ & 1,126.590$^{***}$ \\ 
 & (df = 4; 1985360) & (df = 11; 1985353) & (df = 33; 1985331) & (df = 43; 1985321) \\
\hline 
\hline \\[-1.8ex] 
\textit{Note:}  & \multicolumn{4}{r}{$^{*}$p$<$0.1; $^{**}$p$<$0.05; $^{***}$p$<$0.01} \\ 
\end{tabular} 
\end{table}

Models 3 and 4 include 32-dimensional encoded representations formed by
employing `text-embedding-3-large' on each zip code's name and address
(e.g., `10001 New York, NY'). We chose to use the first 32 dimensions
after empirical testing, as the inclusion of more singular vectors
neither substantially improved model fit nor changed our substantive
results. Our results show that the inclusion of the OpenAI encoding led
to model results that are substantively similar to the inclusion of the
structured data describing geographies, potentially due to
representation being a lower-dimensional capture of the address of each
zip code and therefore an encoding of the unique cultural, economic,
geographic, and social factors of each zip code. Broadly, thus, we show
that the LLM encoding of unstructured data describing the zip codes can
yield similarly useful results as detailed structured data describing
the zip codes---the implicit knowledge graph stored in the LLM's weights
and its expression in the embedding suffices to describe demand in our
application.

Why might a model with the inclusion of 1-digit zip code fixed effects
have a slightly higher R-squared than the model with geographic and
other variables? A detailed look at the regression with 1-digit zip
codes reveals that only four zip code fixed effects corresponding to a
first digit of 2 (DC, MD, NC, SC, VA, and WV), 3 (AL, FL, GA, MS, and
TN), 4 (IN, KY, MI, and OH), and 5 (IA, MN, MT, ND, SD, and WI) are
significant in the regression. The intercept in the regression is
\$38.61, and the coefficient on the coupon is -\$19.54. In contrast, the
zip codes starting with 2, 3, 4, and 5 have coefficients of -\$2.50,
-\$1.47, -\$3.39, and -\$1.73, respectively, which is on the order of
10\% or less of the intercept and 5\% or less of the coupon. Thus, what
we find is that demand across geographies was relatively homogenous for
this firm, with limited variation explained by the location of the
customer---a conclusion that stands across the two stable models using
zip code fixed effects and the projection of the fixed effects to
structured and unstructured data covariates. This is perhaps not
surprising, as this firm was born on the internet, targeted relatively
young consumers, and its products were only retailed direct to consumer.
Therefore, it follows that there was limited variation in revenue per
order across different geographies, with most of the variation explained
by longitude (the extent to which the firm was able to diffuse from the
east to the west) and latitude (the extent to which its products could
be used around the year, including in winter).

\begin{figure}[htbp]
\centering
\begin{subfigure}{.5\textwidth}
  \centering
  \includegraphics[width=\linewidth]{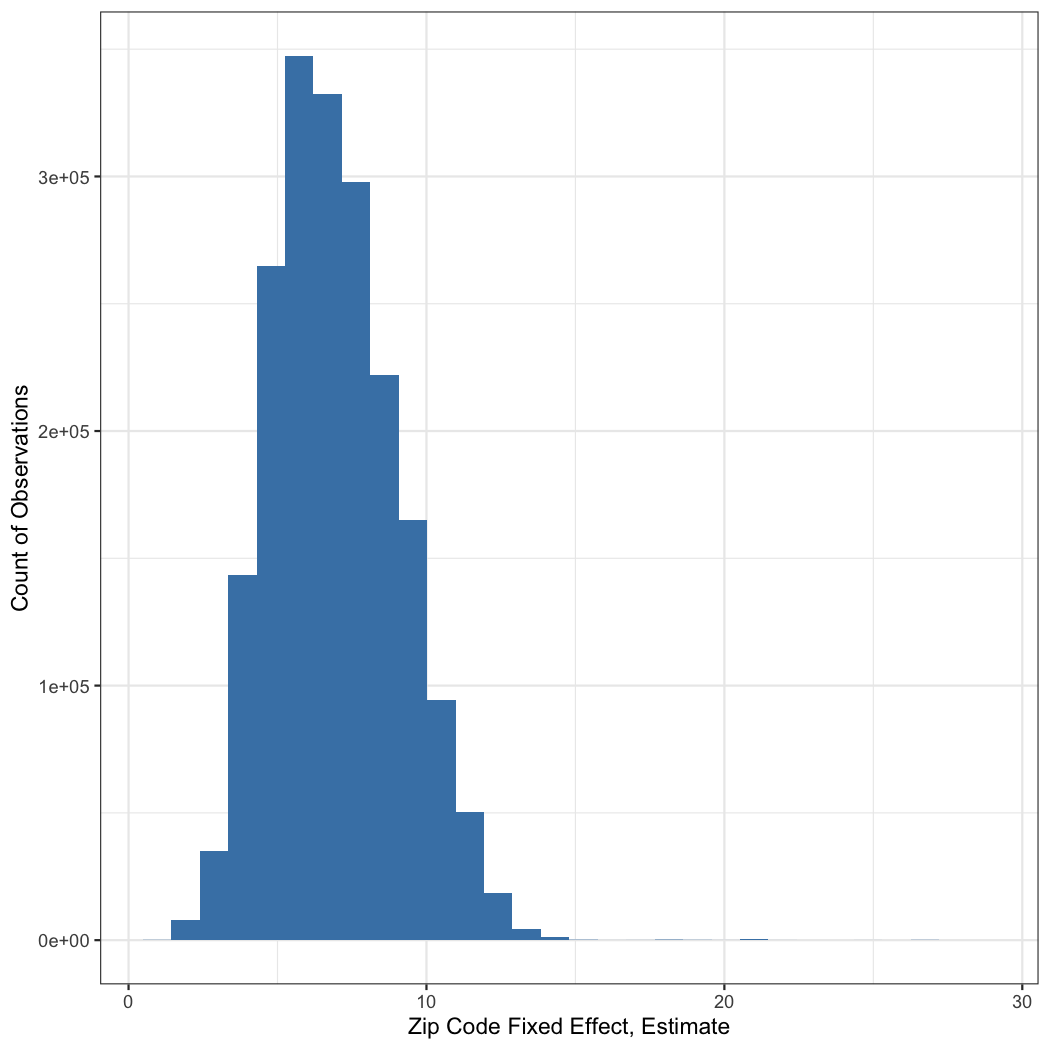}
  \caption{Fixed Effects}
  \label{fig:histogram_fe_a}
\end{subfigure}%
\begin{subfigure}{.5\textwidth}
  \centering
  \includegraphics[width=\linewidth]{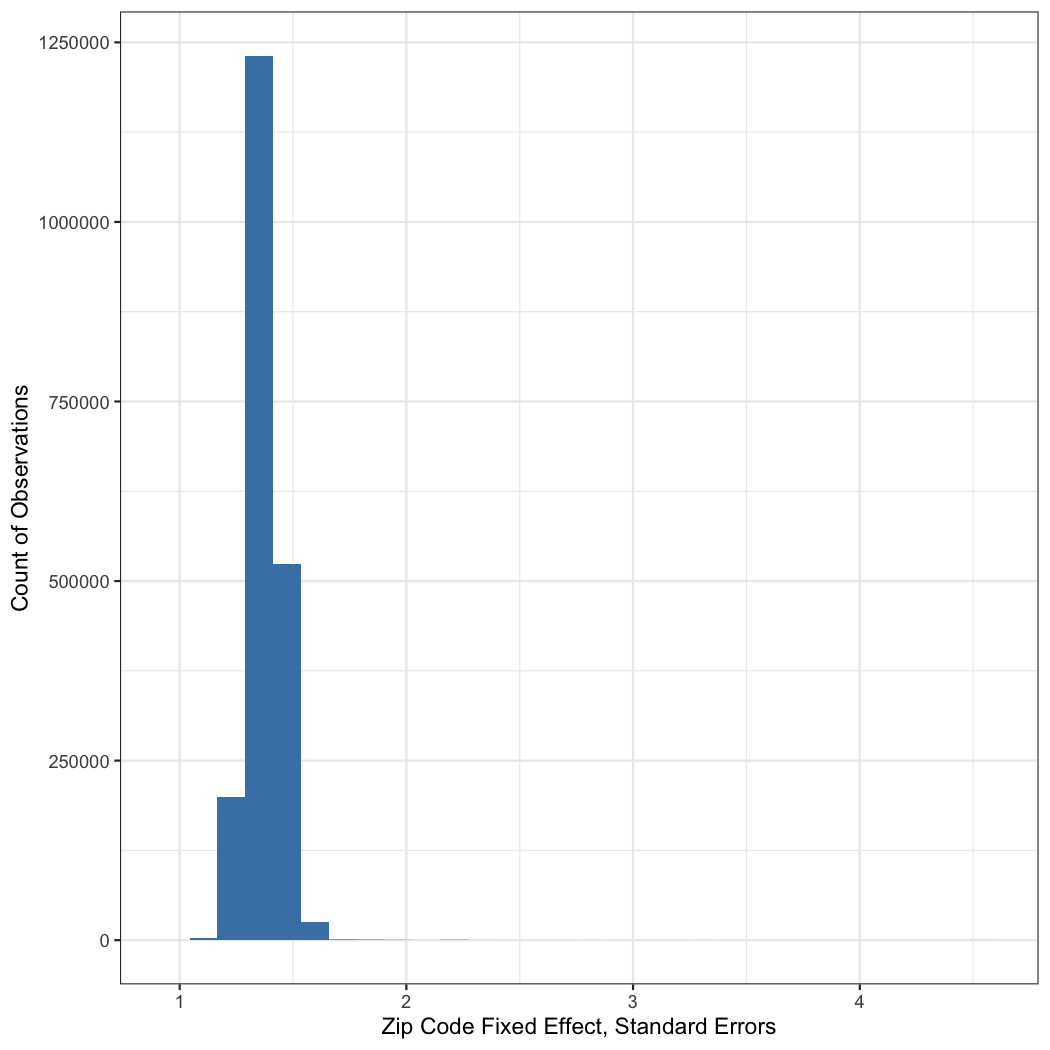}
  \caption{Fixed Effect Standard Errors}
  \label{fig:histogram_fe_b}
\end{subfigure}
\caption{Histograms of Fixed Effects and Standard Errors}
\label{fig:histogram_fe}
\end{figure}

Figure \ref{fig:histogram_fe} presents histograms of the fixed effect
estimates and standard errors from Model 4 in Table
\ref{tab:regressions2}. We see that the estimates range from about 1 to
10, suggesting a difference of about \$10 in spend per order across
geographies (estimation error is centered around \$1 with relatively low
dispersion). This suggests that smoothing out the estimates of spend by
employing structured and unstructured descriptions of variations across
zip codes yields a model that expresses greater geographic differences
in mean spend. The standard errors mostly range from 1.25 to 1.5, with a
mean of 1.374 and a standard deviation of 0.07. 99.22\% of the estimates
are significant against the null at a confidence level of 95\%,
indicating a robust capture of the fixed effects.

\begin{figure}[htbp]
\centering
\includegraphics[width=0.75\linewidth]{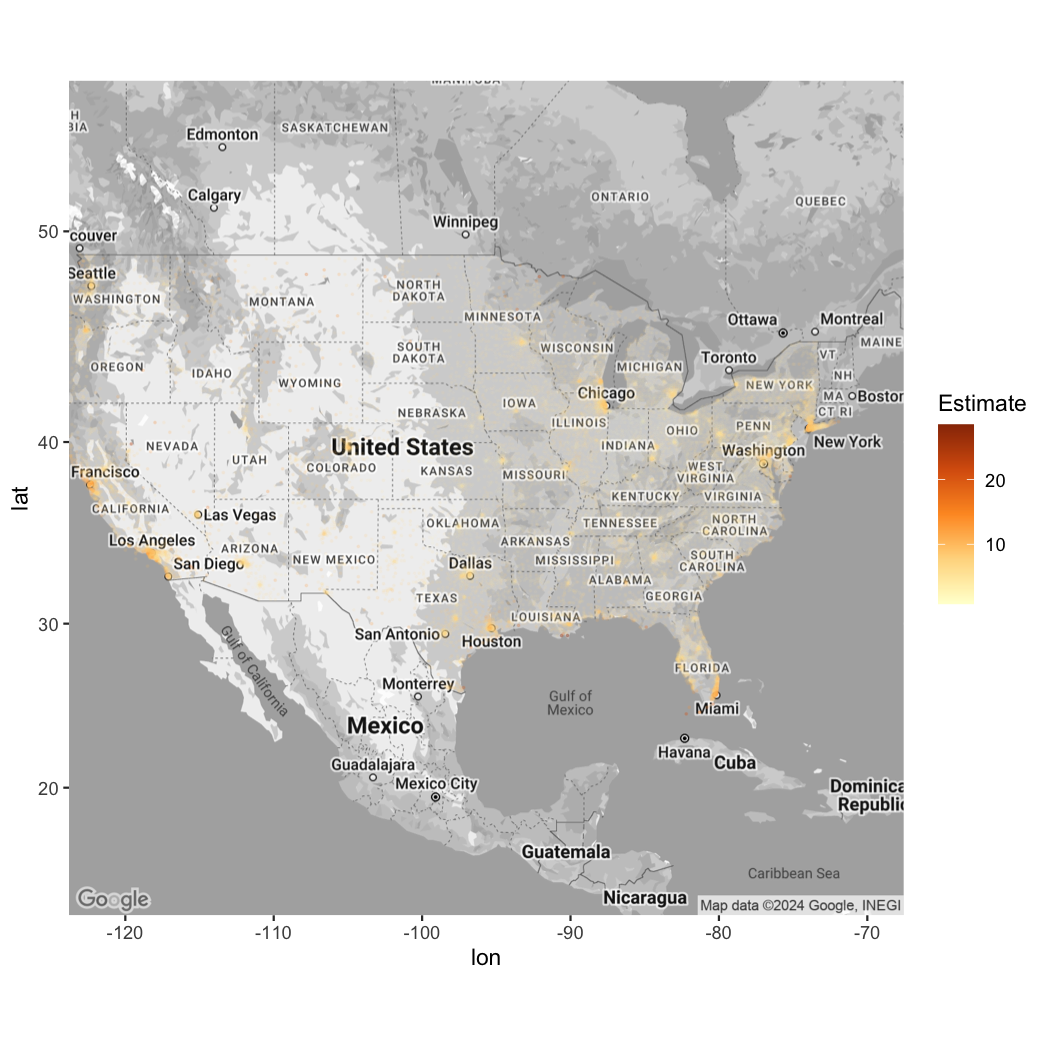}
\caption{Inferred Zip Code Fixed Effects in the Contiguous United States}
\label{fig:map_fe}
\begin{minipage}{\linewidth}
\medskip
\footnotesize
Note: Color and opacity represent fixed effect magnitude.
\end{minipage}
\end{figure}

Figure \ref{fig:map_fe} maps the inferred fixed effects in Model 4 to
the location of the zip code in the United States. We find that revenue
per order, accounting for couponing, varies drastically across the
United States such that in high demand markets (see Figure
\ref{fig:map_field}) not only are more orders placed but also the
revenue per order is higher than in lower demand markets. However, the
effects are clearly not universal. For example, order volume from cities
in the east coast of Florida (e.g.~Miami, FL) is high. However, the
estimated fixed effect and therefore the revenue per order is higher
than other cities with comparable order volume, indicating that these
cities are likely prime targets for offline expansion.

\hypertarget{conclusion}{%
\section{Conclusion}\label{conclusion}}

In this paper, we investigate the integration of categorical variables
with a power-law distribution into causal econometric models. Our
findings indicate that traditional regression estimators, affected by
the sparse and unbounded nature of category levels, may not converge to
accurate values, leading to both inaccurate and imprecise inferences.
This issue arises from the violation of the Donsker conditions,
essential assumptions for the asymptotic normality of estimators and the
parametric complexity in econometric models. The lack of these
conditions undermines the reliability of convergence guarantees.

Through numerical simulations, we reinforce these theoretical insights,
showing that estimates from both the canonical estimator and those using
widely accepted methodological practices---such as excluding or merging
rare categories, or applying regularization techniques like LASSO for
variable selection---yield inconsistent estimates. Moreover, this
problem is exacerbated as models tend to overfit the data, complicating
the reliability of in-sample and asymptotic fit statistics, making these
issues difficult to detect and address.

To address this challenge, we propose a novel estimator that leverages
knowledge of the underlying manifold from which the categorical
variable's expression in the ambient space is derived. In our empirical
work, we position this within a spatial econometrics framework, where a
categorical variable of zip code is distributed as sampled from a
manifold of latitude, longitude, and elevation. We demonstrate that by
using auxiliary information to describe the category levels---whether it
be structured data on the levels or even unstructured data describing
the levels as mapped using a large language model---we can infer
category-level data features that form the scaffolding on which
estimates can be derived. This auxiliary information on category levels
enables our model to surpass the traditional inclusion of dummies in the
canonical model, which is typically considered the efficient estimator.
Unlike the canonical model, which involves a non-parametric capture of
the fixed effects, our approach advocates for an explicit modeling of
these effects.

Our method has broad applicability across various research fields and
industry sectors where high-dimensional categorical data is prevalent.
For example, in healthcare analytics, patient data often includes
categorical variables such as diagnosis codes and treatment codes.
Previous work has shown the potential of using deep learning to generate
unsupervised patient embeddings from electronic health records
(\protect\hyperlink{ref-choi2016doctor}{Choi et al. 2016},
\protect\hyperlink{ref-miotto2016deep}{Miotto et al. 2016}).

Extending our approach could involve using the LLM encoding of each
diagnosis code or treatment code as a set of explanatory covariates for
the fixed effect. This would transform the high-dimensionality
non-parametric model, which might result from including many applicable
diagnosis and treatment codes, into a lower-dimensional, computationally
stable, and tractable model. Such approaches may be particularly
beneficial if data describing the category levels is not only structured
or textual but also multi-modal, allowing for a lower-dimensional
embedding to be derived from visual images, among other sources
(\protect\hyperlink{ref-xu2014deep}{Xu et al. 2014},
\protect\hyperlink{ref-zhang2019deep}{Zhang et al. 2019}). For instance,
images of various diagnoses (e.g., an X-ray image of a fracture) could
serve as inputs to the embedding to bolster its predictive accuracy. We
hope our method and findings prove useful in such endeavors.

\newpage
\singlespacing

\hypertarget{bibliography}{%
\section{Bibliography}\label{bibliography}}

\hypertarget{refs}{}
\begin{CSLReferences}{1}{0}
\leavevmode\vadjust pre{\hypertarget{ref-arellano1991some}{}}%
Arellano M, Bond S (1991) Some tests of specification for panel data:
Monte carlo evidence and an application to employment equations.
\emph{The review of economic studies} 58(2):277--297.

\leavevmode\vadjust pre{\hypertarget{ref-carrizosa2022tree}{}}%
Carrizosa E, Mortensen LH, Morales DR, Sillero-Denamiel MR (2022) The
tree based linear regression model for hierarchical categorical
variables. \emph{Expert Systems with Applications} 203:117423.

\leavevmode\vadjust pre{\hypertarget{ref-center2015future}{}}%
Center PR (2015) The future world religions: Population groth
projections, 2010-2050 why muslim are raising fastest and the
unaffiliated are shirking as share of the world population.
\emph{Diunduh dari https://www. pewforum.
org/2015/04/02/religious-projections-2010-2050}.

\leavevmode\vadjust pre{\hypertarget{ref-cerda2020encoding}{}}%
Cerda P, Varoquaux G (2020) Encoding high-cardinality string categorical
variables. \emph{IEEE Transactions on Knowledge and Data Engineering}
34(3):1164--1176.

\leavevmode\vadjust pre{\hypertarget{ref-chernozhukov2018double}{}}%
Chernozhukov V, Chetverikov D, Demirer M, Duflo E, Hansen C, Newey W,
Robins J (2018) Double/debiased machine learning for treatment and
structural parameters.

\leavevmode\vadjust pre{\hypertarget{ref-choi2016doctor}{}}%
Choi E, Bahadori MT, Schuetz A, Stewart WF, Sun J (2016) Doctor ai:
Predicting clinical events via recurrent neural networks. \emph{Machine
learning for healthcare conference}. (PMLR), 301--318.

\leavevmode\vadjust pre{\hypertarget{ref-darling2005does}{}}%
Darling-Hammond L, Holtzman DJ, Gatlin SJ, Heilig JV (2005) Does teacher
preparation matter? Evidence about teacher certification, teach for
america, and teacher effectiveness. \emph{Education Policy Analysis
Archives/Archivos Analiticos de Politicas Educativas} 13:1--48.

\leavevmode\vadjust pre{\hypertarget{ref-dudley2010universal}{}}%
Dudley RM (2010) Universal donsker classes and metric entropy.
\emph{Selected works of RM dudley}. (Springer), 345--365.

\leavevmode\vadjust pre{\hypertarget{ref-finke2005churching}{}}%
Finke R, Stark R (2005) \emph{The churching of america, 1776-2005:
Winners and losers in our religious economy} (Rutgers University Press).

\leavevmode\vadjust pre{\hypertarget{ref-fox2001religion}{}}%
Fox J (2001) Religion as an overlooked element of international
relations. \emph{International Studies Review} 3(3):53--73.

\leavevmode\vadjust pre{\hypertarget{ref-frana2023demographics}{}}%
Frana PL (2023) Demographics, inc., computerized direct mail, and the
rise of the digital attention economy. \emph{IEEE Annals of the History
of Computing}.

\leavevmode\vadjust pre{\hypertarget{ref-hanushek2016matters}{}}%
Hanushek EA (2016) What matters for student achievement. \emph{Education
next} 16(2):18--26.

\leavevmode\vadjust pre{\hypertarget{ref-hanushek1996aggregation}{}}%
Hanushek EA, Rivkin SG (1996) Aggregation and the estimated effects of
school resources. \emph{Review of Economics \& Statistics} 78(4).

\leavevmode\vadjust pre{\hypertarget{ref-hinz2019separating}{}}%
Hinz J, Hudlet A, Wanner J (2019) Separating the wheat from the chaff:
Fast estimation of GLMs with high-dimensional fixed effects.
\emph{Unpublished Working Paper}.

\leavevmode\vadjust pre{\hypertarget{ref-kane2008does}{}}%
Kane TJ, Rockoff JE, Staiger DO (2008) What does certification tell us
about teacher effectiveness? Evidence from new york city.
\emph{Economics of Education review} 27(6):615--631.

\leavevmode\vadjust pre{\hypertarget{ref-lancaster2000incidental}{}}%
Lancaster T (2000) The incidental parameter problem since 1948.
\emph{Journal of econometrics} 95(2):391--413.

\leavevmode\vadjust pre{\hypertarget{ref-milner2016persistent}{}}%
Milner A, Aitken Z, Kavanagh A, LaMontagne AD, Petrie D (2016)
Persistent and contemporaneous effects of job stressors on mental
health: A study testing multiple analytic approaches across 13 waves of
annually collected cohort data. \emph{Occupational and environmental
medicine} 73(11):787--793.

\leavevmode\vadjust pre{\hypertarget{ref-miotto2016deep}{}}%
Miotto R, Li L, Kidd BA, Dudley JT (2016) Deep patient: An unsupervised
representation to predict the future of patients from the electronic
health records. \emph{Scientific reports} 6(1):1--10.

\leavevmode\vadjust pre{\hypertarget{ref-rivkin2005teachers}{}}%
Rivkin SG, Hanushek EA, Kain JF (2005) Teachers, schools, and academic
achievement. \emph{econometrica} 73(2):417--458.

\leavevmode\vadjust pre{\hypertarget{ref-smith1990classifying}{}}%
Smith TW (1990) Classifying protestant denominations. \emph{Review of
Religious Research} 31(3):225--245.

\leavevmode\vadjust pre{\hypertarget{ref-steensland2000measure}{}}%
Steensland B, Robinson LD, Wilcox WB, Park JZ, Regnerus MD, Woodberry RD
(2000) The measure of american religion: Toward improving the state of
the art. \emph{Social forces} 79(1):291--318.

\leavevmode\vadjust pre{\hypertarget{ref-suits1957use}{}}%
Suits DB (1957) Use of dummy variables in regression equations.
\emph{Journal of the American Statistical Association} 52(280):548--551.

\leavevmode\vadjust pre{\hypertarget{ref-tutz2016regularized}{}}%
Tutz G, Gertheiss J (2016) Regularized regression for categorical data.
\emph{Statistical Modelling} 16(3):161--200.

\leavevmode\vadjust pre{\hypertarget{ref-vershynin2018high}{}}%
Vershynin R (2018) \emph{High-dimensional probability: An introduction
with applications in data science} (Cambridge university press).

\leavevmode\vadjust pre{\hypertarget{ref-voas2013modernization}{}}%
Voas D, McAndrew S, Storm I (2013) Modernization and the gender gap in
religiosity: Evidence from cross-national european surveys.
\emph{K{ö}lner Zeitschrift F{ü}r Soziologie \& Sozialpsychologie} 65(1).

\leavevmode\vadjust pre{\hypertarget{ref-woodberry2012missionary}{}}%
Woodberry RD (2012) The missionary roots of liberal democracy.
\emph{American political science review} 106(2):244--274.

\leavevmode\vadjust pre{\hypertarget{ref-woodberry2012measure}{}}%
Woodberry RD, Park JZ, Kellstedt LA, Regnerus MD, Steensland B (2012)
The measure of american religious traditions: Theoretical and
measurement considerations. \emph{Social Forces} 91(1):65--73.

\leavevmode\vadjust pre{\hypertarget{ref-wooldridge2010econometric}{}}%
Wooldridge JM (2010) \emph{Econometric analysis of cross section and
panel data} (MIT press).

\leavevmode\vadjust pre{\hypertarget{ref-xu2014deep}{}}%
Xu Y, Mo T, Feng Q, Zhong P, Lai M, Eric I, Chang C (2014) Deep learning
of feature representation with multiple instance learning for medical
image analysis. \emph{2014 IEEE international conference on acoustics,
speech and signal processing (ICASSP)}. (IEEE), 1626--1630.

\leavevmode\vadjust pre{\hypertarget{ref-zhang2019deep}{}}%
Zhang Z, Zhao Y, Liao X, Shi W, Li K, Zou Q, Peng S (2019) Deep learning
in omics: A survey and guideline. \emph{Briefings in functional
genomics} 18(1):41--57.

\end{CSLReferences}

\end{document}